\documentclass[twocolumn,showpacs,preprintnumbers,nofootinbib,prd,
superscriptaddress,10pt]{revtex4-1}

\usepackage{lmodern}
\usepackage{amsmath,amssymb}
\usepackage[normalem]{ulem}
\usepackage{textcomp}
\usepackage{hyperref}
\usepackage{bm}
\usepackage{graphicx}
\usepackage{psfrag}
\usepackage{aas_macros}
\usepackage[usenames,dvipsnames]{xcolor}
\usepackage[utf8]{inputenc}
\usepackage{lipsum} 
\usepackage{multirow}

\definecolor{darkgreen}{rgb}{0,0.5,0}

\hypersetup{
    bookmarks=true,         % show bookmarks bar?
    unicode=false,          % non-Latin characters in Acrobat???s bookmarks
    pdftoolbar=true,        % show Acrobat???s toolbar?
    pdfmenubar=true,        % show Acrobat???s menu?
    pdffitwindow=false,     % window fit to page when opened
    pdfstartview={FitH},    % fits the width of the page to the window
	pdftitle={Blind spots and biases: the dangers of ignoring eccentricity in GW signals from BBHs},
    pdfauthor={Divyajyoti et al.},
    pdfsubject={Gravitational Waves},   % subject of the document
    pdfkeywords={gravitational waves, eccentric compact binaries, general relativity},
    pdfnewwindow=true,      % links in new window
    colorlinks=true,       % false: boxed links; true: colored links
    linkcolor=red,          % color of internal links
    citecolor=Blue,        % color of links to bibliography
    filecolor=magenta,      % color of file links
    urlcolor=Blue,           % color of external links
    linktocpage=true
}

\allowdisplaybreaks[4]

\DeclareFontFamily{OT1}{pzc}{}
\DeclareFontShape{OT1}{pzc}{m}{it}{<-> s * [1.10] pzcmi7t}{}
\DeclareMathAlphabet{\mathpzc}{OT1}{pzc}{m}{it}

\begin{document}

\title{Blind spots and biases: The dangers of ignoring eccentricity in gravitational-wave signals from binary black holes}

\date{\today}

\author{Divyajyoti}
\email{divyajyoti.physics@gmail.com}
\affiliation{Department of Physics, Indian Institute of Technology Madras, Chennai 600036, India}
\affiliation{Centre for Strings, Gravitation and Cosmology, Department of Physics, Indian Institute of Technology Madras, Chennai 600036, India}

\author{Sumit Kumar}
\email{sumit.kumar@aei.mpg.de}
\affiliation{Max-Planck-Institut f{\"u}r Gravitationsphysik (Albert-Einstein-Institut), D-30167 Hannover, Germany}
\affiliation{Leibniz Universit{\"a}t Hannover, D-30167 Hannover, Germany}

\author{Snehal Tibrewal}
\email{snehaltibrewal@gmail.com}
\affiliation{Department of Physics, Indian Institute of Technology Madras, Chennai 600036, India}
\affiliation{Centre for Strings, Gravitation and Cosmology, Department of Physics, Indian Institute of Technology Madras, Chennai 600036, India}

\author{Isobel M. Romero-Shaw}
\email{ir346@cam.ac.uk}
\affiliation{Department of Applied Mathematics and Theoretical Physics, Cambridge CB3 0WA, United Kingdom}
\affiliation{Kavli Institute for Cosmology Cambridge, Madingley Road, Cambridge CB3 0HA, United Kingdom}

\author{Chandra Kant Mishra}
\email{ckm@iitm.ac.in}
\affiliation{Department of Physics, Indian Institute of Technology Madras, Chennai 600036, India}
\affiliation{Centre for Strings, Gravitation and Cosmology, Department of Physics, Indian Institute of Technology Madras, Chennai 600036, India}

\begin{abstract}
Most gravitational wave (GW) events observed so far by the LIGO and Virgo detectors are consistent with mergers of binary black holes (BBHs) on quasi-circular orbits. However, some events, such as GW190521, are also consistent with having non-zero orbital eccentricity at detection, which can indicate that the binary formed via dynamical interactions. Active GW search pipelines employing quasi-circular waveform templates are inefficient for detecting eccentric mergers. Additionally, analysis of GW signals from eccentric BBH with waveform models neglecting eccentricity can lead to biases in the recovered parameters. Here, we explore the detectability and characterisation of eccentric signals when searches and analyses rely on quasi-circular waveform models. We find that for a reference eccentric population, the fraction of events having fitting factor (FF) $<0.95$ can be up to $\approx2.2\%$ compared to $\approx0.4\%$ for the baseline population. This results in the loss in signal recovery fraction for up to $6\%$ for the region in parameter space with non-negligible eccentricity ($e_{10} > 0.01$) and high mass ratio ($q > 3$). We perform parameter estimation (PE) for non-spinning and aligned-spin eccentric injections of GWs from binaries of total mass $M=35$~M$_\odot$, based on numerical relativity simulations and an EOB based inspiral-merger-ringdown model (\textsc{TEOBResumS}), and recover them using both quasi-circular and eccentric waveform models. For $e_{20}\sim 0.1$, analyses using quasi-circular waveform models are unable to recover the injected chirp mass within the $90\%$ credible interval. Further, for these low-mass injections, spin-induced precession \textit{does not} mimic eccentricity, with PE correctly ruling out high values of effective precession parameter $\chi_p$. For injections of $e_{20}\sim 0.1$, PE conducted with an inspiral-only eccentric waveform model correctly characterises the injected signal to within $90\%$ confidence, and recovers the injected eccentricities, suggesting that such models are sufficient for characterisation of low-mass eccentric BBH. 

\end{abstract}

\date{\today}

\maketitle
%%%
\section{Introduction}
\label{sec:intro}

Since the first detection of gravitational waves (GWs)~\cite{Abbott:2016blz}, the LIGO-Virgo-KAGRA (LVK) collaboration has reported about 85 GW candidates from binary black hole (BBH) mergers \cite{LIGOScientific:2018mvr, LIGOScientific:2020ibl, LIGOScientific:2021usb, KAGRA:2021vkt}. While most of the detected signals are consistent with GW emission from inspiralling BBHs on quasi-circular orbits, several events have been argued to be more consistent with coming from binaries with non-negligible orbital eccentricity at detection \citep[e.g.,][]{Romero-Shaw:2021ual, Romero-Shaw:2022xko, Iglesias:2022xfc}. In particular, GW190521~\citep{Abbott:2020mjq, Abbott:2020tfl} has been interpreted as coming from a moderately to highly eccentric BBH \citep{Romero-Shaw:2020thy, Gamba:2021gap, Gayathri:2020coq}, in addition to other interpretations such as merger of Proca stars \citep{CalderonBustillo:2021:GW190521}, and merger of a binary black hole system with high spin-precession \cite{Miller:2023ncs}. Non-negligible orbital eccentricity measured at detection in the LVK sensitive frequency range (above $10$~Hz) implies that the radiating BBH was driven to merge by external influences: for example, as part of a field triple \citep[e.g.,][]{Antonini:2017:triples}, in a densely populated star cluster \citep[e.g.,][]{Rodriguez:2018:GCs}, or in the accretion disk of a supermassive black hole \citep[e.g.,][]{Tagawa:2021:AGN}.
\\
\\
Search pipelines based on matched-filtering methods use quasi-circular waveform templates, motivated by the expected efficient circularisation via GW emission of compact binary orbits during the late stages of their evolution~\cite{PhysRev.136.B1224}.
However, binaries formed through dynamical processes in dense stellar environments \cite{PortegiesZwart:1999nm, 2010MNRAS.407.1946D, Rodriguez:2016kxx, Banerjee:2017mgr, DiCarlo:2019pmf, Mapelli2020, Mandel:2018hfr, Samsing:2017xmd, Fragione:2018yrb, Samsing:2020tda} or through Kozai-Lidov processes \cite{Kozai:1962zz, Lidov62} in field triples \cite{Martinez:2020lzt, 2016ARA&A..54..441N}, may be observed in ground based detectors such as advanced LIGO \citep{KAGRA:2013rdx, LIGOScientific:2014pky, 2020PhRvD.102f2003B} and Virgo \citep{VIRGO:2014yos,2019PhRvL.123w1108A} with residual eccentricities $\gtrsim0.1$ \citep[e.g.,][]{2011A&A...527A..70K, Lower:2018seu, Antonini:2017:triples, Samsing:2017rat, Rodriguez:2018:GCs, Zevin:2021rtf, Tagawa:2021:AGN}. 
While pipelines employing quasi-circular templates should be able to detect the majority of systems with eccentricities $e_{10}\lesssim 0.1$ at a GW frequency of $10$ Hz \cite{Favata:2021vhw} if observed with current LIGO-Virgo detectors, binaries with larger eccentricities would require constructing template banks for matched-filter searches including the effect of eccentricity \citep[e.g.,][]{Brown:2009ng, Zevin:2021rtf}. Moreover, the presence of even small eccentricities ($e_{10}\sim0.01-0.05$) can induce bias in extracting source properties~\cite{Favata:2013rwa, Abbott:2016wiq, Favata:2021vhw, Ramos-Buades:2019uvh, Narayan:2023vhm, Guo:2022ehk, Saini:2022igm, Bhat:2022amc, Bonino:2023:eccentricity, GilChoi:2022nhs}. As existing detectors upgrade to their LIGO A+/Voyager \cite{McClelland:T1500290-v3, LIGO:2020xsf} configurations, improving their low-frequency sensitivity, neglecting eccentricity in detection and parameter estimation pipelines may lead to incorrect inference of source properties and/or failure to identify the presence of eccentric signals in data. This problem is likely to be exacerbated in detections made with next-generation ground-based instruments such as Cosmic Explorer \cite{Evans:2021gyd, LIGOScientific:2016wof, Reitze:2019iox} and the Einstein Telescope \citep{Punturo_2010, Hild:2010id, ET-0106C-10}, since their sensitivity to frequencies $\sim 1$~Hz and above should enable them to frequently observe systems with detectable eccentricities~\cite{Lower:2018seu, Chen:2020lzc}.

In searches for compact binary coalescence signals, computation time and availability of waveform models play crucial roles. In recent years, there have been some targeted searches for eccentric systems \cite{Cheeseboro:2021rey, Pal:2023dyg, Lenon:2021zac, Cokelaer:2009hj, Wang:2021qsu, Ravichandran:2023qma, Ramos-Buades:2020eju, LIGOScientific:2019dag, LIGOScientific:2023lpe, Taylor:2015kpa, Martel:1999tm, Tai:2014bfa}, and upper limits have been provided in the absence of detection: with data from first two observing runs of LIGO and Virgo detector network, Nitz \textit{et al} (2019) \citep{Nitz:2019spj} provided $90\%$ credible upper limits for binary neutron stars as $\sim 1700$ mergers Gpc$^{-3}$ Yr$^{-1}$, for eccentricities $\leq 0.43$, for dominant mode frequency at $10$~Hz. For sub-solar mass binaries, Nitz \textit{et al} (2021) \citep{Nitz:2021vqh} provided $90\%$ credible upper limits for $0.5-0.5$ ($1.0-1.0$)~M$_\odot$ binaries to be $7100$ ($1200$) Gpc$^{-3}$ Yr$^{-1}$. Dhurkunde \textit{et al} (2023) \citep{Dhurkunde:2023qoe} searched for aligned spin neutron star binaries in the O3 public data of Advanced LIGO and Virgo and constrained the local merger rate with $90\%$ upper limits to be $\leq 150$ Gpc$^{-3}$ Yr$^{-1}$ for binary neutron star systems in the field, and $\leq 100$ Gpc$^{-3}$ Yr$^{-1}$ for neutron star-black hole binaries from various dynamical channels.

While inspiral-only models for GW signals from eccentric compact binary systems are sufficiently accurate to NR simulations of inspirals, and are rapid enough to generate for use in direct parameter estimation via Bayesian inference \cite{Konigsdorffer:2006zt, Yunes:2009yz, Klein:2010ti, Mishra:2015bqa, Moore:2016qxz, Tanay:2016zog, Klein:2018ybm, Boetzel:2019nfw, Ebersold:2019kdc, Moore:2019xkm, Klein:2021jtd, Khalil:2021txt, Paul:2022xfy, Henry:2023tka}, their use may be limited to low mass eccentric (typically $\lesssim25M_{\odot}$)~\cite{LIGOScientific:2011jth} binaries due to the absence of the merger and ringdown in the signal model. 
Waveform models containing the inspiral, merger, and ringdown (IMR) are under development and/or available for use \citep[e.g.,][]{Hinder:2017sxy, Huerta:2017kez, Cao:2017ndf, Chiaramello:2020ehz, Islam:2021mha, Chattaraj:2022tay, Ramos-Buades:2021adz, Liu:2023dgl}; these are generally slower to generate than their quasi-circular counterparts, and Bayesian inference using these models has usually required reduction of accuracy conditions \citep[e.g.,][]{OShea:2021faf}, using likelihood reweighting techniques \citep[e.g.,][]{Romero-Shaw:2019itr}, or utilising highly computationally expensive parallel inference on supercomputer clusters \citep[e.g.,][]{pBilby, Romero-Shaw:2022xko}. Additionally, of all available eccentric waveform models, there is only one that includes \textit{both} of the parameters required to fully describe an eccentric orbit \citep{Ramos-Buades:2023yhy}: the orbital eccentricity at a reference epoch, and a parameter describing the orientation of the ellipse at the same epoch, e.g., the mean anomaly. Neglecting the second eccentricity parameter has the potential to lead to additional biases in source parameter recovery \citep{Clarke:2022fma, Ramos-Buades:2023yhy}.
\\
\\
Methods currently in use for eccentric \textit{searches} \cite{Klimenko:2005xv,Tiwari:2015gal, Coughlin:2015jka, Lower:2018seu, Salemi:2019uea}, with little or no dependence on signal model, are sensitive to high masses \citep[ideally $\gtrsim 70M_{\odot}$][]{LIGOScientific:2019dag,LIGOScientific:2023lpe}. Nevertheless, one can infer the presence of orbital eccentricity in signals found by standard searches tuned to quasi-circular BBH by employing available eccentric waveform models via Bayesian parameter estimation \citep[e.g.,][]{Lower:2018seu, Romero-Shaw:2019itr, Romero-Shaw:2020thy, Romero-Shaw:2020:GW190425, Lenon:2020:eccBNS, Romero-Shaw:2021ual, Romero-Shaw:2022xko, Romero-Shaw:2022fbf, Iglesias:2022xfc, Gamba:2021gap, Bonino:2023:eccentricity} or by comparing the data directly to numerical relativity waveform simulations of GW from eccentric BBH \citep{Gayathri:2020coq}. A caveat to all of the Bayesian inference studies mentioned above is the absence of spin-induced precession \cite{PhysRevD.49.6274} in the waveform model employed: since both eccentricity and misaligned spins introduce modulations to the gravitational waveform \cite{OShea:2021faf, Romero-Shaw:2020thy}, it may be critical to account for both spin precession and eccentricity while aiming to measure either or both of the two effects, particularly for GWs from high-mass BBHs \citep{Romero-Shaw:2022fbf, Ramos-Buades:2019uvh}.\footnote{While the model in \citep{Klein:2021jtd} does include the effect of spin-precession and eccentricity, the equations of motion derived there are targeted toward binaries on eccentric orbits undergoing spin-induced precession that can efficiently be integrated on the radiation-reaction timescale, such as stellar-origin black hole binaries observed by LISA \citep{Babak:2021mhe}. The applicability of this model for LIGO sources may be explored in a future study.
}
\\
\\
Even though the currently available eccentric waveform models may not include the effect of spin-induced precession, these are still useful for studying systematic errors incurred due to the neglect of orbital eccentricity in waveform models used in LVK catalogs \citep[e.g.,][]{KAGRA:2021vkt}. Eccentric versions of the effective-one-body (EOB) waveforms \cite{Liu:2019jpg, Hinderer:2017jcs, Albanesi:2022xge} including higher modes \cite{Ramos-Buades:2021adz, Nagar:2021gss, Iglesias:2022xfc, Liu:2021pkr} and an eccentric numerical relativity (NR) surrogate model \cite{Islam:2021mha, Yun:2021jnh, Huerta:2017kez} are available. In this work, we focus on assessing the impact of orbital eccentricity in the dominant quadrupole mode ($\ell=2, |m|=2$) of the waveform, which we argue is a reasonable representation of the GW emission for the majority of observed source types, which are equal-mass and low-spin.\footnote{While this may not be true generically, the fact that only two of the events observed so far show the presence of a higher order mode \cite{Abbott:2020khf, LIGOScientific:2020stg}
and that most detections are consistent with a zero-effective spin scenario~\cite{LIGOScientific:2020ibl} favours the assumption.} A summary of the investigations presented here is included below.  
\subsection{Summary of analysis and results}
\label{sec:summary}

We assess the impact of employing a circular template bank for GW searches when the source population may exhibit eccentricity. To achieve this, we simulate diverse source populations, covering the parameters of binary black hole masses and eccentricity. We determine the detection efficiency of the population by comparing the inherent signal strength to the values obtained from the circular template bank. We also investigate the regions in the parameter space where the largest loss in signal strength occurs due to the difference between injection and recovery waveform model or due to non-inclusion of eccentricity.

\vspace{1em}

In order to explore the effect of eccentricity on various inferred parameters of a BBH system, we perform parameter estimation (PE) on injected non-spinning and aligned spin eccentric signals generated using numerical relativity (NR) codes, and recover them using different waveform models with different combinations of spins and eccentricity. Our aim is to observe and quantify the biases incurred due to absence of eccentricity in the recovery waveform when analyzing eccentric signals, and to verify that those biases can be corrected to a certain extent by using the currently available eccentric waveforms.
\\
\\
We perform multiple sets of zero-noise NR hybrid and waveform approximant injections which include synthetic GWs consistent with non-spinning and aligned-spin BBHs in eccentric orbits, and recover these injections with a variety of either quasi-circular or eccentric dominant mode ($\ell=2, |m|=2$) waveform models including no spin, aligned spins, or misaligned (precessing) spins.\footnote{We have also explored the simulated noise case for one of the injection sets. Those results are discussed in Appendix \ref{appendix:noisy_injs}.} 
Due to a lack of available waveform models, we are unable to investigate injection recovery using models with both spin-precession and eccentricity. We observe that the state-of-the-art inspiral-merger-ringdown (IMR) quasi-circular waveforms of the {\textsc{PhenomX} \citep{Pratten:2020fqn, Pratten:2020ceb} family do not recover the true chirp mass of an eccentric signal.
When a non-spinning eccentric injection is analysed using quasi-circular waveforms, the chirp mass posterior is biased from the true value. The magnitude of this bias is same regardless of the spin configuration (non-spinning, spin-aligned, or spin-precessing) used in the recovery waveform. Moreover, the spin posteriors for eccentric injections do not show any additional bias compared to the quasi-circular injections. 
Since the posteriors for spin-precession parameter ($\rm{\chi_p}$) peak correctly at 0 for non-spinning injections, therefore, we conclude that for low-mass (total mass of $35$~M$_\odot$), long-inspiral, non-spinning, eccentric binary systems, parameter estimation with a precessing-spin waveform model \textbf{will not lead to a false positive value of precession}. This supports the conclusions drawn in \citet{Romero-Shaw:2022fbf}, where the authors demonstrate that eccentricity and spin-precession are distinguishable for signals with more than a few cycles in-band. 
\\
\\
Further, analyzing these same signals using a computationally efficient inspiral-only eccentric waveform results in a significant reduction of the biases in the posteriors on the intrinsic parameters of the binary, leading to the recovery of the true values within the $90\%$ credible bounds. This implies that for GWs from low-mass and aligned- or low-spin BBHs, \textbf{inspiral-only waveforms that are readily available within the \texttt{LALSuite} framework are adequate for accurate recovery of the source parameters.} This is true despite the neglect of the second eccentric parameter, e.g. the mean anomaly, in the recovery waveform. We also see clear correlation between chirp mass and eccentricity posteriors for both the non-spinning and aligned spin eccentric injections, in agreement with the previous studies \citep{Favata:2021vhw, Ramos-Buades:2019uvh, Ramos-Buades:2023yhy, Bonino:2023:eccentricity}, in addition to a mild correlation of eccentricity with the effective inspiral spin parameter $\chi_\text{eff}$ [see Eq. \eqref{eq:chi_eff}] when the injection is aligned-spin and eccentric, consistent with the correlations seen in \citet{OShea:2021faf}.\\
\\
While we see that the posteriors obtained from inspiral-only eccentric models are able to recover the injected values of parameters (as opposed to IMR quasi-circular models which yield biased results), it is likely that biases can be further reduced and posteriors be better constrained if an IMR eccentric waveform model including the second eccentric parameter is used for recovery. However, our results demonstrate that even eccentric waveform models with limited physics (no merger or ringdown, no spin-precession, no higher modes, no mean anomaly) can reduce errors in the inference of BBH parameters, and may therefore be a useful stepping stone toward analysis with full IMR models including eccentricity, spin-precession, and higher modes. \\
\\
This paper is organized as follows. In Sec.~\ref{sec:gw_searches}, we describe the methodology used for quantifying the sensitivity reduction of a GW search to eccentric signals, along with other metrics such as fitting factor and signal recovery fraction. We present our results for the detectability of eccentric systems in Sec.~\ref{subsec:detection}. In the second half of the paper, we focus on parameter estimation of eccentric binaries, starting in Sec.~\ref{sec:pe}. We start with non-spinning eccentric systems (Sec.~\ref{subsec:non-spin-inj}) and then move to aligned spin eccentric systems in Sec.~\ref{subsec:align-spin-inj}. Finally, in Sec.~\ref{sec:concl}, we include a comprehensive summary of our findings and future directions.

\section{Blind spots: Reduced GW search sensitivity to eccentric binaries}
\label{sec:gw_searches}

\begin{figure*}[ht!]
    \centering
    \includegraphics[width=0.48\linewidth]{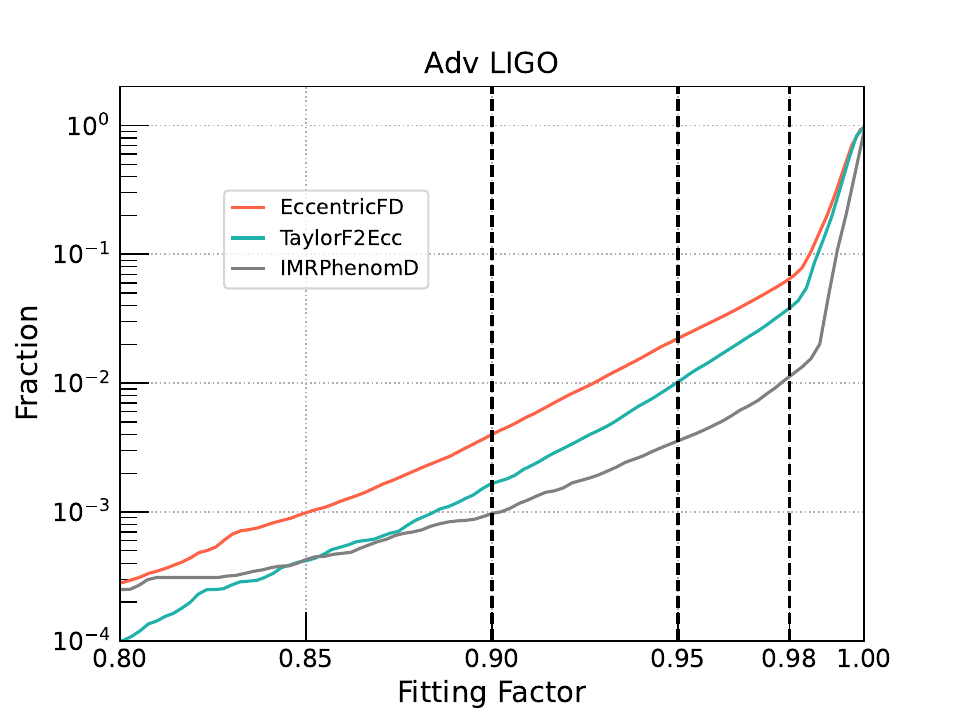}
    \includegraphics[width=0.48\linewidth]{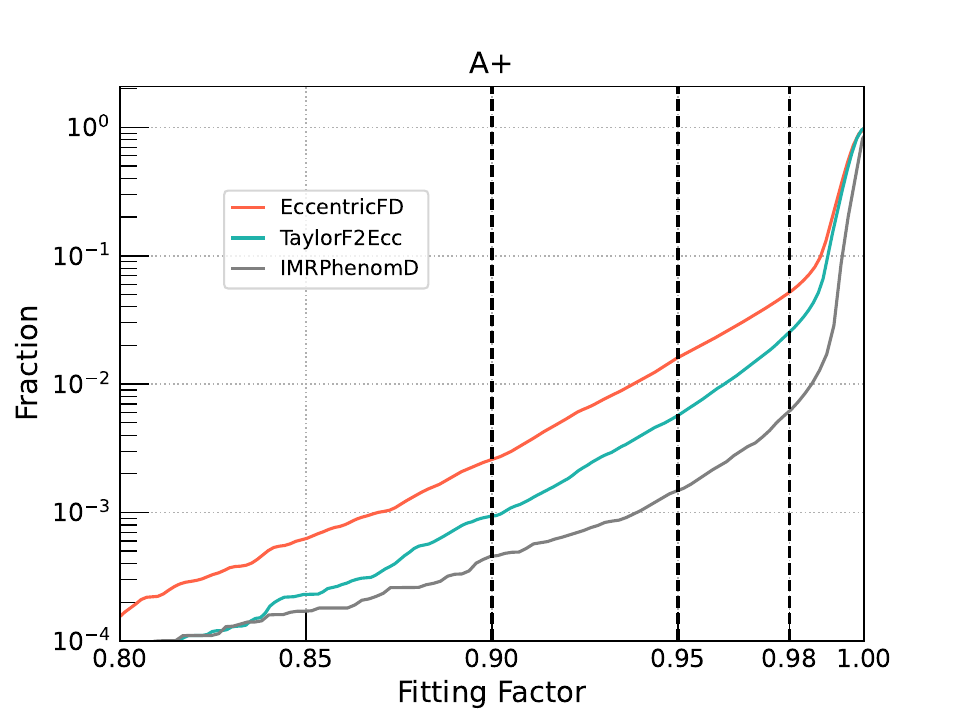}

    \caption{Cumulative fraction of events above a given fitting factor $FF$ for various populations, distributed uniformly in masses and log-uniformly in eccentricity (measured at $10$~Hz) with the match calculated against the standard template bank, shown here for detector sensitivities: Adv-LIGO (left)  and $A+$ (right). The grey curves show the fraction recovered for the reference population with no eccentricity, while the green and red curves show the fraction recovered for the eccentric population represented by the \textsc{TaylorF2Ecc} and \textsc{EccentricFD} models, respectively. Three vertical dashed black lines show the fitting factor values of $0.9$, $0.95$, and $0.98$ increasing in value from the left. These plots show that if we use the quasi-circular template bank to search for the population which contains a log-uniform distribution of eccentricities, we fail to detect a higher fraction of signals in searches. E.g. in the left panel (Adv LIGO sensitivity): eccentric population constructed with \textsc{TaylorF2Ecc} (\textsc{EccentricFD}) waveform has $\approx$ 1 percent ($\approx$ 2.2 percent) events with fitting factor less than 0.95. $FF$ for baseline model \textsc{IMRPhenomD} is $\approx$ 0.4 percent. In the right panel ($A+$): eccentric population constructed with \textsc{TaylorF2Ecc} (\textsc{EccentricFD}) waveform has $\approx$ 0.6 percent ($\approx$ 1.6 percent) events with $FF$ less than 0.95. $FF$ for baseline model \textsc{IMRPhenomD} is $\approx$ 0.1 percent.}
    \label{fig:cumulative_histogram_ff}
\end{figure*}

\begin{figure*}[ht!]
    \centering
    \includegraphics[width=0.48\linewidth]{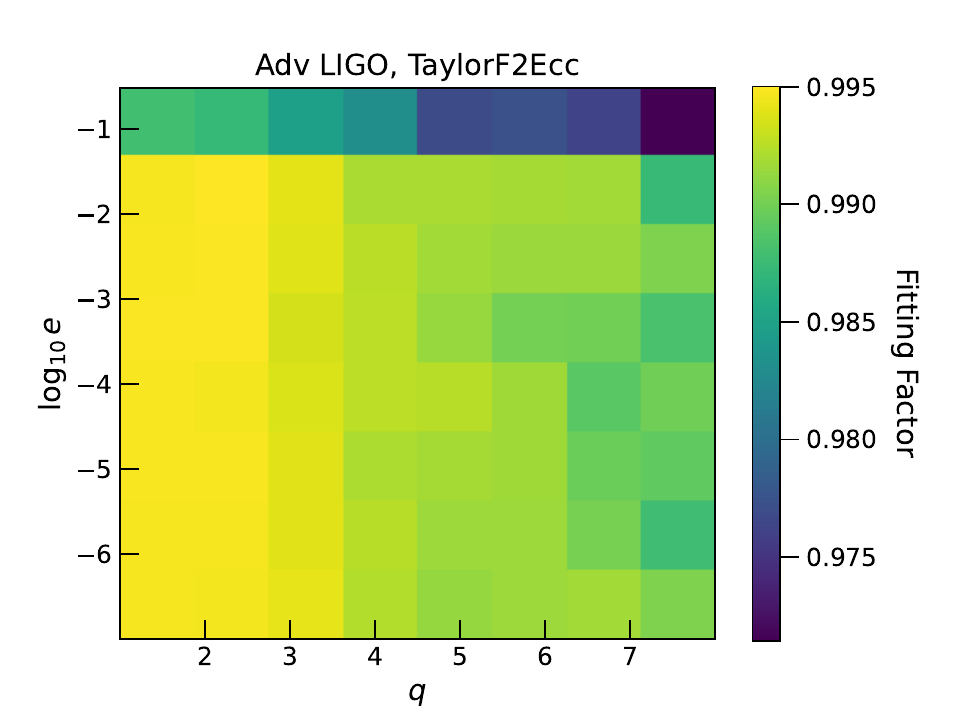}
    \includegraphics[width=0.48\linewidth]{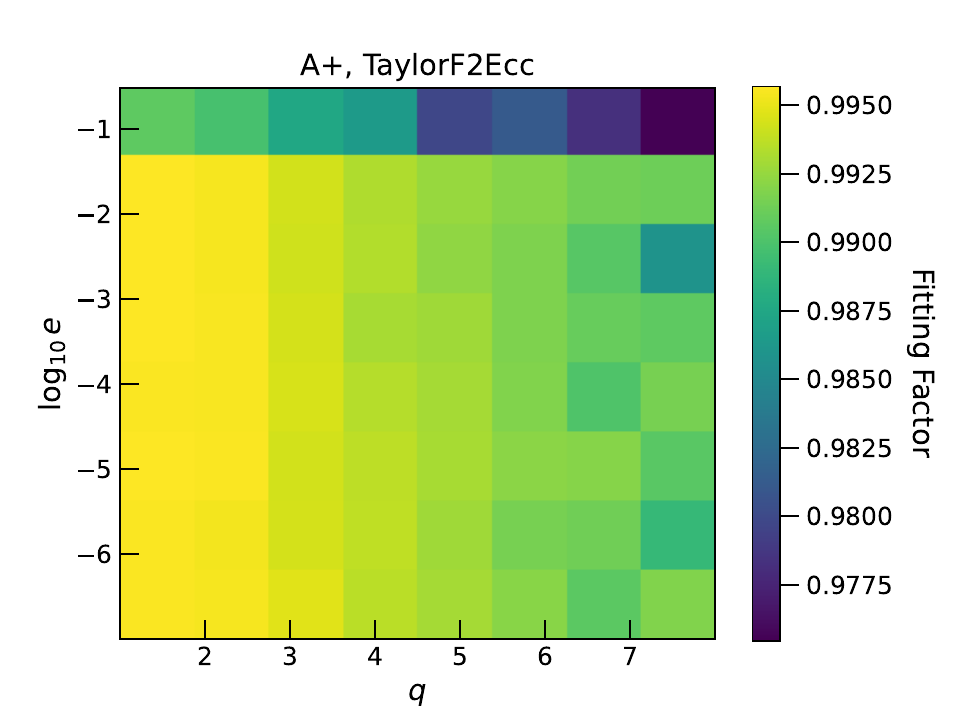}
    \includegraphics[width=0.48\linewidth]{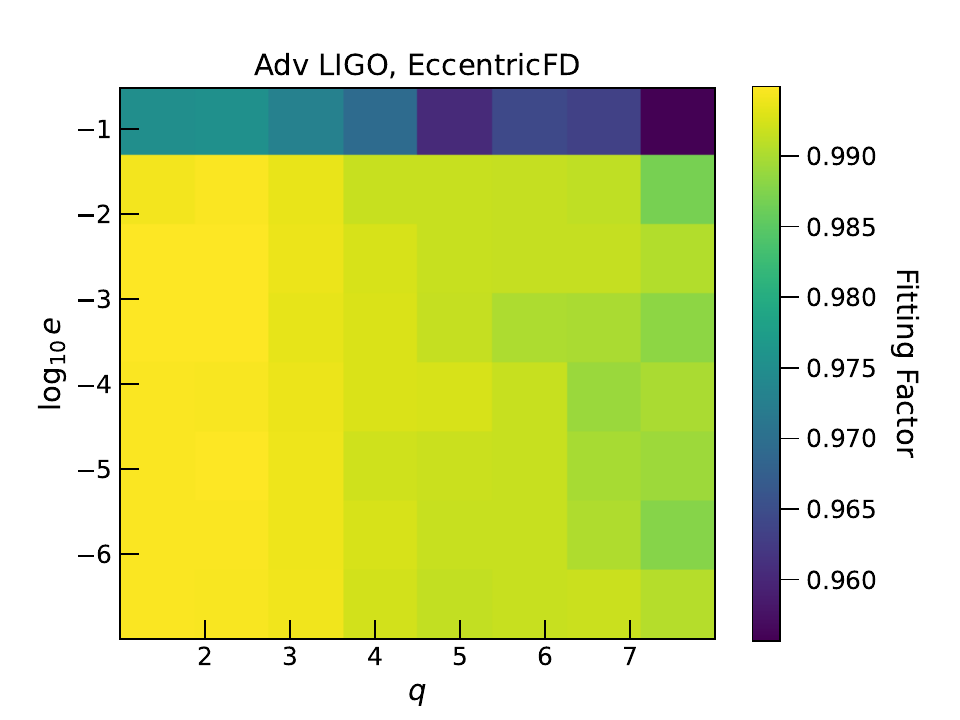}
    \includegraphics[width=0.48\linewidth]{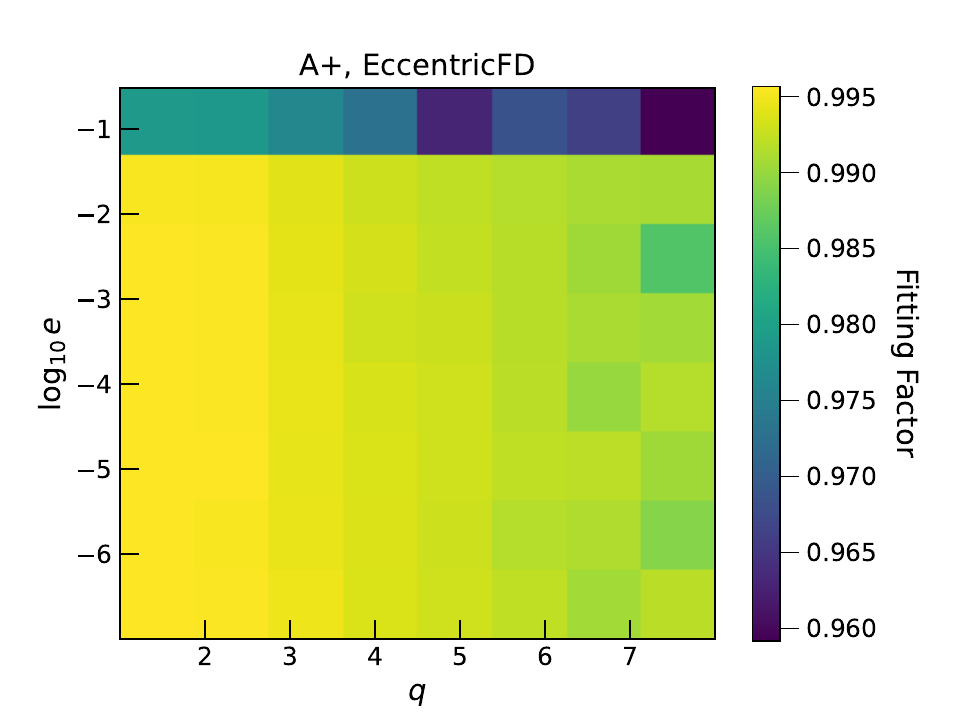}
    \caption{The Fitting Factor (FF) varying with mass ratio $q$ and $\log_{10}{(e)}$ for a population uniform in component masses and log-uniform in eccentricity $e_{10}$ measured at 10 Hz. Top row represents the population generated with \textsc{TaylorF2Ecc} and bottom row represents the population generated with \textsc{EccentricFD}. For all plots, we use the recovery template bank generated with \textsc{IMRPhenomD}. The left column shows results assuming the detector sensitivity of advLIGO and the right column shows results assuming the detector sensitivity of $A+$. The maximal loss in fitting factor occurs for high mass ratios ($q=m_1/m_2$) and high eccentricity regimes, with high eccentricity values playing dominating role. We use only the inspiral part of the waveform, up to frequency corresponding to innermost stable circular orbit (ISCO), to calculate the $FF$.}
    \label{fig:ff_ecc_q}
\end{figure*}

In this section, we investigate the fraction of events that might be missed by searches due to neglecting eccentricity in template banks used for matched-filter based searches such as \texttt{PyCBC} \cite{Usman:2018imj}, or \texttt{GSTLAL} \cite{2017PhRvD..95d2001M}. The GW data represented by time series $s(t)$ contains noise $n(t)$, and may contain signal $g(t)$. The signal model $h(t, \theta_i)$ is dependent on parameters $\theta_i$, which define the intrinsic properties of the binary from which the GW emanates. The modeled searches perform matched filtering between signal templates $\Tilde{h}$ and the detector data $\Tilde{s}$ in the Fourier domain. The correlation of data $\Tilde{s}(f)$ with the signal model $\Tilde{h}(f)$ weighted by the power spectral density (PSD) $S_n(f)$ of the noise is given by,
\begin{equation}
    <s|h> = 4\int_0^\infty \frac{\Tilde{s}(f)\Tilde{h}(f)}{S_n(f)}
\end{equation}

The GW searches use discrete template banks. Such searches will always miss a fraction of signals because i) templates may not be accurate representations of the real signal, especially if these signals include additional physics (e.g., eccentricity, misaligned spins, higher modes) not included in the template bank and ii) discreteness of the bank. In order to quantify the loss of sensitivity of the search, we define the match between the template $h(\theta_i, \Phi)$ and a signal $g$ in terms of the overlap between them as:
\begin{equation}
    m(g,h(\theta_i)) =  \text{max}(<g|h(\theta_i, \Phi)>), \label{eqn:match}
\end{equation}
where $h(\theta_i, \Phi)$ is a template with intrinsic parameters $\theta_i$ and extrinsic parameters $\Phi$. The right-hand side (rhs) in equation \eqref{eqn:match} is maximized over all the extrinsic parameters. The fitting factor $FF(g)$, for a signal $g$, is defined as \citep{Balasubramanian:1994uy, Apostolatos:1995pj, Dhurkunde:2022aek}:
\begin{equation}
    FF(g) = \text{max}(m(g, h(\theta_i))), \label{eqn:fitting_factor}
\end{equation}
where rhs in Eq.~\eqref{eqn:fitting_factor} is maximized over all the templates $h(\theta_i)$. For a given signal g, the match, as defined in Eq.~\eqref{eqn:match}, is estimated with all the templates in the bank, and the maximum match is reported as FF for that signal. Therefore, FF tells us the maximum match an injected signal can obtain for a template bank in a region of parameter space. Estimating FF is an excellent tool to realize which part of the parameter space of injections we get the best/worst recovery in quantifiable terms.

FF can vary across the parameter space $\theta_i$. Therefore, for a population of signals, in order to get fraction of recovered signals relative to an optimal search, for which FF=1, we use the metric described in \cite{Buonanno:2002fy}, which takes into account the intrinsic SNR of the signal to calculate the signal recovery fraction (SRF), defined as \citep{Dhurkunde:2022aek}:
\begin{equation}
    SRF \equiv \frac{\sum_{i=0}^{n_s - 1} \text{FF}^3_\text{TB}(s_i) \sigma^3(s_i)}{\sum_{i=0}^{n_s-1}\sigma^3(s_i)} ,\label{eqn:signal_fraction_recovery}
\end{equation}
\noindent
where FF$_\text{TB}$ is the fitting factor for a volumetric distribution of $n_s$ sources using a template bank and $\sigma(s_i)$ is the intrinsic loudness of each signal $s_i$. We calculate the SRF for a given distribution of sources and a template bank.

\subsection{Reduced detectability of eccentric systems}
\label{subsec:detection}

We use a reference population distributed uniformly in source frame masses between $m^{\rm source}_{1,2} \in [5, 50]$ and redshift up to 3. For eccentricity parameter, log-uniform distribution is used while the sources are distributed uniformly in sky. In this section, we consider non-spinning population to quantify the effects of eccentricity parameter.
The population is generated with three different waveform models: 
\begin{itemize}
    \item \textbf{No eccentricity distribution:} We use the \textsc{IMRPhenomD} \cite{Husa:2015iqa, Khan:2015jqa} waveform model to inject non-spinning, quasi-circular signals and use a template bank with the same waveform for recovery. This serves as the optimal search and we expect the maximal recovery of the injected signals.
    \item \textbf{With eccentricity distribution:} For these sets of injections, we use two waveform models: we generate one set with \textsc{TaylorF2Ecc} \cite{Moore:2016qxz, Kim:2019abc}, and the other with \textsc{EccentricFD} \cite{Huerta:2014eca}. We use the same component masses as above, and a log-uniform eccentricity distribution\footnote{The choice of the eccentric parameter is motivated by the expected realistic population of eccentric binaries \citep{Zevin:2021rtf}, and since most of the eccentric binaries are expected to have low eccentricity, we want to resolve the parameter space in that regime. However, for the FF estimation, we expect that the choice of the reference population will not affect the conclusions drawn if the injected population is dense enough.} in the range ${e_{10}} \in [10^{-7},0.3]$ defined at $10$ Hz. We choose this upper limit to stay within the regions of validity of the waveform models.
\end{itemize}

For the recovery, we use the ``quasi-circular'' template bank (non-spinning) constructed with the waveform model \textsc{IMRPhenomD}. We use stochastic placing algorithms \cite{PhysRevD.80.104014, Babak:2008rb} implemented in \texttt{PyCBC} to generate template bank for component masses in range $\in [3, 200]~M_{\odot}$, using the minimal match criteria of 0.98. We use this template bank to quantify the fraction of lost signals if the intrinsic population has some intrinsic eccentricity distribution. We calculate the optimal SNR for each injection using the template bank and then estimate the $FF$ for the set of injections. We use the low frequency cutoff of $10$~Hz and detector sensitivities for i) advanced LIGO \cite{PyCBC-PSD:aLIGO}, and ii) $A+$ design sensitivity \cite{PyCBC-PSD:Aplus}. In Fig.~\ref{fig:cumulative_histogram_ff}, we show the fitting factors for all three injection sets considered above. As expected, the quasi-circular injection set generated with \textsc{IMRPhenomD} and recovered with the quasi-circular template bank gives us the maximum $FF$. The upper cut-off frequency, for each system, is chosen to be the frequency corresponding to the innermost stable circular orbit ($f_\text{ISCO}$) for a test particle orbiting a Schwarzschild black hole. We use the minimum low frequency of $10$~Hz to calculate the match via Eq.~\eqref{eqn:match}. The loss of $FF$ is visible with both eccentric injection sets. In Fig.~\ref{fig:ff_ecc_q}, we explore the parameter regime where the reduction in $FF$ is maximum. As expected, larger eccentricity values ($e_{10}>0.01$) give us lower $FF$. We also notice that the combination of high mass ratio ($q=m_1/m_2; m_1>m_2$) and high eccentricity gives us maximum loss in the $FF$. We propose that more extreme mass ratios lead to a larger reduction in $FF$ for the same value of $e_{10}$ because binaries with more extreme $q$ have longer GW signals in-band, and therefore have more inspiral cycles over which the mismatch due to eccentricity accumulates. 
\\
\\
\begin{table*}[t]
\begin{tabular}{|c|c|c|c|c|}
\hline
  \textbf{\begin{tabular}[c]{@{}c@{}}Injection Waveform\end{tabular}} &
  \hspace{5pt}$\textbf{\begin{tabular}[c]{@{}c@{}}SRF \\ Full Range \end{tabular}}$\hspace{5pt} &
  \hspace{5pt}$\textbf{\begin{tabular}[c]{@{}c@{}}SRF \\ ($\mathbf{e_{10}>0.01}$) \end{tabular}}$\hspace{5pt} &
  \hspace{5pt}$\textbf{\begin{tabular}[c]{@{}c@{}}SRF \\ ($\mathbf{q>3}$) \end{tabular}}$\hspace{5pt} & \hspace{5pt}$\textbf{\begin{tabular}[c]{@{}c@{}}SRF \\ ($\mathbf{e_{10}>0.01}$) \& ($\mathbf{q>3}$) \end{tabular}}$\hspace{5pt} \\ \hline
IMRPhenomD     & 0.992 (0.992) & -     & 0.986 (0.997) & -  \\ \hline
TaylorF2Ecc & 0.989 (0.987) & 0.973 (0.97) & 0.923 (0.978) & 0.923 (0.948)   \\ \hline
EccentricFD & 0.989 (0.987) & 0.969 (0.963)   &  0.923 (0.979)  & 0.918 (0.944) \\ \hline
\end{tabular}
\caption{The SRF, as described in Eq.~\eqref{eqn:signal_fraction_recovery}, is calculated for the injection sets described in the text to quantify the reduced detectability of eccentric signals when circular template bank is used. For recovery, we use a template bank designed for non-eccentric searches using \textsc{IMRPhenomD} waveform model. For each column, two numbers are shown: one for Advanced LIGO search sensitivity and the numbers in the bracket are quoted for A+ search sensitivity. The SRF for the optimal search (injection with \textsc{IMRPhenomD}) indicates the maximum. For eccentric injections, the loss in the SRF is maximum in the parameter space ($e_{10}>0.01, q>3$) which is affected most due to loss in FF.}
\label{table:SRF}
\end{table*}

Figures \ref{fig:cumulative_histogram_ff} and \ref{fig:ff_ecc_q} show that for a given population of eccentric signals, there will be loss of $FF$ for signals with eccentricity $e_{10} > 0.01$. As mentioned earlier, the loss of FF indicates the loss of sensitivity of the search. For instance, in Fig.~\ref{fig:cumulative_histogram_ff}, it can be seen that, for AdvLIGO search sensitivity, $\sim99\%$ of quasi-circular binaries in the population are recovered with FF $> 0.98$. In contrast, for eccentric binary systems simulated using \textsc{TaylorF2Ecc} (\textsc{EccentricFD}), this percentage is $\sim96\%$ ($\sim94\%$) when both of these populations are being recovered using a quasi-circular template bank. For A+ sensitivity, $\sim99\%$ of quasi-circular binaries in the population are recovered with the same threshold of FF, whereas for eccentric binary systems simulated using \textsc{TaylorF2Ecc} (\textsc{EccentricFD}), $\sim97\%$ ($\sim95\%$) of injections are recovered above the same threshold, when using a quasi-circular template bank. This indicates that a quasi-circular template bank employed for searches will lead to losing a fraction of events when the population consists of binaries in eccentric orbits. This trend becomes more prominent for more extreme values of mass ratio $q$. The extent of the overall search volume loss depends on the proportion of high-eccentricity signals in the population. In order to include eccentricity in GW searches, we require (i) efficient eccentric waveform models and (ii) a low computational cost in comparison to the gain in the search volume. 

The SRF depends on the intrinsic source population under consideration. If the fraction of signals with high eccentricity ($>0.01$) is large, we expect to fail to recover a higher fraction of them. In order to estimate SRF, we choose a network of three detectors: HLV, with two LIGO detectors H and L at Hanford and Livingston respectively, and the Virgo (V) detector in Italy. For the full population described above, we estimate SRF to be 0.992 for optimal search (injection and recovery done with \textsc{IMRPhenomD}) for both the Advanced LIGO and A+ detector sensitivities. We kept the same sensitivity for Virgo \citep{PyCBC-PSD:AdvVirgo} in both the networks. With eccentric injections using \textsc{EccentricFD}, the SRF for full population is estimated to be 0.989 (0.987) for Advanced LIGO (A+) detector sensitivity. For another set of eccentric injections generated using TaylorF2Ecc, the estimated SRF is  0.989 (0.987) for Advanced LIGO (A+) detector sensitivity. This indicates that the presence of eccentricity in GW signal reduces the overall SRF if non-eccentric recovery models are used. Moreover if the population has a significant number of events from the parameter space which is responsible for most loss in FF, the SRF is further reduced, indicating failure of recovering comparatively large fraction of events in that parameter region. For a targeted region in parameter space of non-negligible eccentricity ($e_{10}>0.01$) and high mass ratio ($q>3$), we summarize the results in Table \ref{table:SRF}. In this targeted region, we can get the value of SRF as low as $\sim$0.918 compared to the SRF of $\sim$0.99 for optimal pipeline.

To gain insights into a realistic population, we create another injection set. This set incorporates a power-law distribution of source frame masses, consistent with GWTC-3 population analysis \cite{KAGRA:2021duu}, and an eccentricity distribution drawn from simulations outlined in \cite{Kremer:2019iul, Zevin:2021rtf}. We limit the source frame mass distribution to the range [5, 50]$M_{\odot}$, aligning with the template banks we generated. We use the same HLV detector network with two sensitivities for LIGO detectors. For this injection set, the SRF for the baseline model was calculated at $\sim$0.986 for both Advanced LIGO and A+ sensitivity. Focusing on targeted regions ($e_{10}>0.01, q>3$), the SRF is found to be $\sim$0.944 (0.946) for \textsc{EccentricFD} (\textsc{TaylorF2Ecc}) injections with Advanced LIGO design sensitivity, and $\sim$0.927 (0.928) for \textsc{EccentricFD} (\textsc{TaylorF2Ecc}) injections with A+ design sensitivity.

While we might detect eccentric signals via either quasi-circular template-based searches \textit{or} unmodeled searches, increasing the true fraction of the underlying eccentric population that will enter into our catalogs, bias could be introduced in the subsequently inferred parameters. For this reason, quantifying the bias introduced by analysing eccentric signals with quasi-circular waveform models is necessary. We turn our attention to this in the following section.

\section{Biases: Mischaracterizing eccentric binaries with parameter estimation}
\label{sec:pe}

In this section, we present results from injection analyses. We assess waveform systematics due to the neglect of eccentricity in PE studies, employing quasi-circular waveform models to recover injections into detectors with zero noise (i.e., the detector response to the signal is accounted for, but no additional Gaussian noise is added to the power spectral density representing the detector's sensitivity). We also perform injections into simulated Gaussian noise and analyze those injections too, for completeness; the results, which are consistent with those with zero noise, and are presented in Appendix~\ref{appendix:noisy_injs}. We perform two sets of injections: one with non-spinning simulations based on NR \cite{Chattaraj:2022tay, Hinder:2017sxy}, and one using an EOB-based IMR signal model, \textsc{TEOBResumS} \cite{Nagar:2018zoe, Chiaramello:2020ehz, Mora:2002gf, Nagar:2021xnh, Placidi:2021rkh, Albanesi:2022ywx, Albanesi:2022xge, Placidi:2023ofj}, for aligned-spin injections. For non-spinning injections, we analyse quasi-circular and eccentric signals with mass ratios $q=(m_1/m_2)=(1, 2, 3)$ and a fixed total mass of $M=35 M_\odot$. For $q=1$ injection, since the mass ratio prior is restricted to $q \geq 1$, the posterior almost entirely lies above the injected value, skewing the posteriors for other correlated parameters (this has been discussed in detail in the following sections). Hence, for aligned-spin eccentric injection, we choose $q=(1.25, 2, 3)$ with the same total mass, and drop $q=1$ case. We employ state-of-the-art quasi-circular waveform models (with and without spins) to recover the injections via Bayesian parameter estimation (PE). We also perform PE with an approximate inspiral eccentric waveform \textsc{TaylorF2Ecc}~\cite{Moore:2016qxz, Kim:2019abc}. The approximate eccentric model used here for PE does not include contributions from spin corrections associated with eccentricity and is based on a Taylor approximant~\cite{Damour:2002kr, Buonanno:2009zt} different from the one used in \textsc{TEOBResumS}. We assume that our sources are at a distance of $410$~Mpc and inclined at an arbitrary angle of $30^\circ$ to the line of sight. The right-ascension ($\alpha$), declination ($\delta$), and polarization ($\psi$) angles are chosen arbitrarily with the values $\sim 164^\circ$, $60^\circ$, and $60^\circ$ respectively, and the geocent time ($t_\text{gps}$) is taken to be 1137283217~s. Since the SNR of a GW signal depends on extrinsic parameters in addition to intrinsic parameters like mass and eccentricity, different extrinsic parameters may lead to different SNRs, changing the widths of the posteriors presented here. 
\\
\\
The Bayesian posterior probability for a parameter $\Vec{\theta}$, given the data $\Vec{s}$ and a GW model $h$, is given by

\begin{equation}
  p(\Vec{\theta}|\Vec{s},h) = \frac{p(\Vec{s}|\Vec{\theta},h) p(\Vec{\theta},h)}{p(\Vec{s})}\,,
\end{equation}
where $p(\Vec{s}|\Vec{\theta},h)$ represents the likelihood, $p(\Vec{\theta})$ is the prior, and $p(\Vec{s}|h)$ represents the evidence. We also calculate Bayes factors between recoveries with eccentric and quasi-circular models, defined as:

\begin{equation}
    \mathcal{B}_\text{E/C} = \frac{p(\Vec{s}|h_1)}{p(\Vec{s}|h_2)}
\end{equation}

where $E$ and $C$ correspond to eccentric and quasi-circular recoveries respectively, and $h_i$ enumerates the waveform approximants under consideration.\footnote{For more information about the method, see Ref. \cite{Biwer:2018osg}.} To estimate parameters, we use the \texttt{PyCBC Inference Toolkit}~\cite{Biwer:2018osg} and explore the parameter space that includes chirp mass ($\mathcal{M}$), mass ratio ($q$), time of coalescence ($t_c$), luminosity distance ($d_L$), phase of coalescence ($\phi_c$), inclination angle ($\iota$), right ascension($\alpha$), and declination($\delta$). For aligned spin recoveries, we use two additional parameters corresponding to the $z$-components of the spin vectors \textit{viz.}~$\left({\chi_{\rm 1z}}~\&~{\chi_{\rm 2z}}\right)$. For recoveries with spin-precession, we use isotropic spin distribution sampling the six spin components in spherical polar coordinates \textit{viz.} the spin magnitudes ($a_i$) and the spin angles ($S_i^\Theta$, $S_i^\Phi$).\footnote{where $i=[1,2]$ corresponds to the binary components, and $\Theta$ and $\Phi$ indicate the polar and azimuthal angles respectively used in spherical polar coordinate system.} For recoveries with spins, we also obtain posteriors on two additional spin parameters. The first is effective spin parameter, $\chi_\text{eff}$, that captures the spin effects along the direction of the angular momentum axis and is defined as \cite{Ajith:2009bn, Santamaria:2010yb}:

\begin{equation}
    \chi_\text{eff} = \frac{m_1 \chi_\text{1z} + m_2 \chi_\text{2z}}{m_1 + m_2}, 
\label{eq:chi_eff}
\end{equation}

where $\chi_\text{1z}$ and $\chi_\text{2z}$ are the components of the two spin vectors in the direction of the angular momentum vector. The second parameter is spin-precession parameter, $\rm{\chi_p}$, that measures the spin effects in the orbital plane of the binary, and is defined in terms of the perpendicular spin vectors, $S_{i\perp}=|\hat{L}\times (\vec{S_i}\times \hat{L})|$, where $\Vec{S}_i$ is the individual spin angular momentum vector
of the compact object in the binary with mass $m_i$, and $\hat{L}$ represents the unit vector along the angular momentum axis of the binary. The effective spin-precession parameter can be written as \cite{Schmidt:2012rh, Hannam:2013oca, Schmidt:2014iyl}:

\begin{equation}
    {\chi_\mathrm{p}}=\frac{1}{A_{1}m_{1}^2} \max(A_{1}S_{1\perp}, A_{2} S_{2\perp}),
\end{equation}
where, $A_{1}=2+(3/2q)$ and $A_{2}=2+(3q/2)$ are mass parameters defined in terms of the mass ratio $q=m_1/m_2>1$.}
For eccentric recovery, we include an additional eccentricity ($e$) parameter in the parameter space.\footnote{Refer Table \ref{table:priors} in Appendix \ref{appendix:priors} for complete information on priors.} 
In our analysis, we marginalize over the polarization angle. The following subsections provide details of the specific injections, as well as various variations of recovery-waveform spin settings with which these injections have been recovered. While discussing the results in the following subsections, we make use of the term ``recovery" to indicate a result in which the $90\%$ credible interval of the posterior includes the injected value, and the systematic bias (difference between the median value and injected value) in the posterior is less than the width of the posterior (at $90\%$ confidence). We judge that the result shows a significant bias if the injected value lies completely outside the $90\%$ credible interval of the posterior. As indicated earlier, these biases are dependent on SNRs which, in this study, fall in the range of typical SNRs observed in the GW event catalogs. We use the HLV network with design sensitivities of Advanced LIGO \cite{PyCBC-PSD:aLIGO} and Virgo \cite{PyCBC-PSD:AdvVirgo} detectors to perform all the parameter estimation analyses shown here.

\subsection{Non-spinning, eccentric injections}
\label{subsec:non-spin-inj}

\begin{table}[t]
\begin{tabular}{|c|c|c|c|c|}
\hline
\textbf{S. No.} &
  \textbf{\begin{tabular}[c]{@{}c@{}}Injection Simulation ID / \\ Waveform\end{tabular}} &
  \hspace{5pt}$\mathbf{q}$\hspace{5pt} &
  \textbf{$\mathbf{e_{20}}$} &
  \textbf{$\mathbf{\chi_\text{eff}}$} \\ \hline
1  & SXS:BBH:1132     & 1 & 0     & -   \\ \hline
2  & HYB:SXS:BBH:1355 & 1 & 0.104 & -   \\ \hline
3  & HYB:SXS:BBH:1167 & 2 & 0.0   & -   \\ \hline
4  & HYB:SXS:BBH:1364 & 2 & 0.104 & -   \\ \hline
5  & HYB:SXS:BBH:1221 & 3 & 0.0   & -   \\ \hline
6  & HYB:SXS:BBH:1371 & 3 & 0.123 & -   \\ \hline
7  & \textsc{TEOBResumS}       & 1.25 & 0.0   & 0.3 \\ \hline
8  & \textsc{TEOBResumS}       & 1.25 & 0.1   & 0.3 \\ \hline
9  & \textsc{TEOBResumS}       & 2 & 0.0   & 0.3 \\ \hline
10  & \textsc{TEOBResumS}       & 2 & 0.1   & 0.3 \\ \hline
11  & \textsc{TEOBResumS}       & 3 & 0.0   & 0.3 \\ \hline
12 & \textsc{TEOBResumS}       & 3 & 0.1   & 0.3 \\ \hline
\end{tabular}
\caption{List of non-spinning, eccentric NR hybrid simulations (constructed in Ref.~\cite{Chattaraj:2022tay}) and injections based on aligned-spin eccentric EOB model \textsc{TEOBResumS} \cite{Nagar:2018zoe} used as injections. Columns include a unique hybrid ID for each simulation (SXS IDs are retained for identification with SXS simulations \citep{Boyle:2019kee, Buchman:2012dw, Chu:2009md, Hemberger:2013hsa, Scheel:2014ina, Blackman:2015pia, SXS:catalog, Lovelace:2011nu, Lovelace:2010ne, Mroue:2013xna, Mroue:2012kv, Lovelace:2014twa, Kumar:2015tha, Lovelace:2016uwp, Abbott:2016nmj, Hinder:2013oqa, Abbott:2016apu, Chu:2015kft, Abbott:2016wiq, Varma:2018mmi, Varma:2018aht, Varma:2019csw, Varma:2020bon, Islam:2021mha, Ma:2021znq} used in constructing the hybrids) and the name of the waveform model used for generating injections, information concerning the mass ratio ($q=m_1/m_2$), eccentricity ($e_{20}$) at the reference frequency of $20$~Hz for a total mass of $M=35$ M$_\odot$, and effective spin $\chi_\text{eff}$ defined in Eq.~\eqref{eq:chi_eff} (only shown for spinning injections).} 
\label{table:hybrids}
\end{table}

\begin{figure*}[t!]
    \centering

    \includegraphics[trim=100 10 100 30, clip, width=0.715\linewidth]{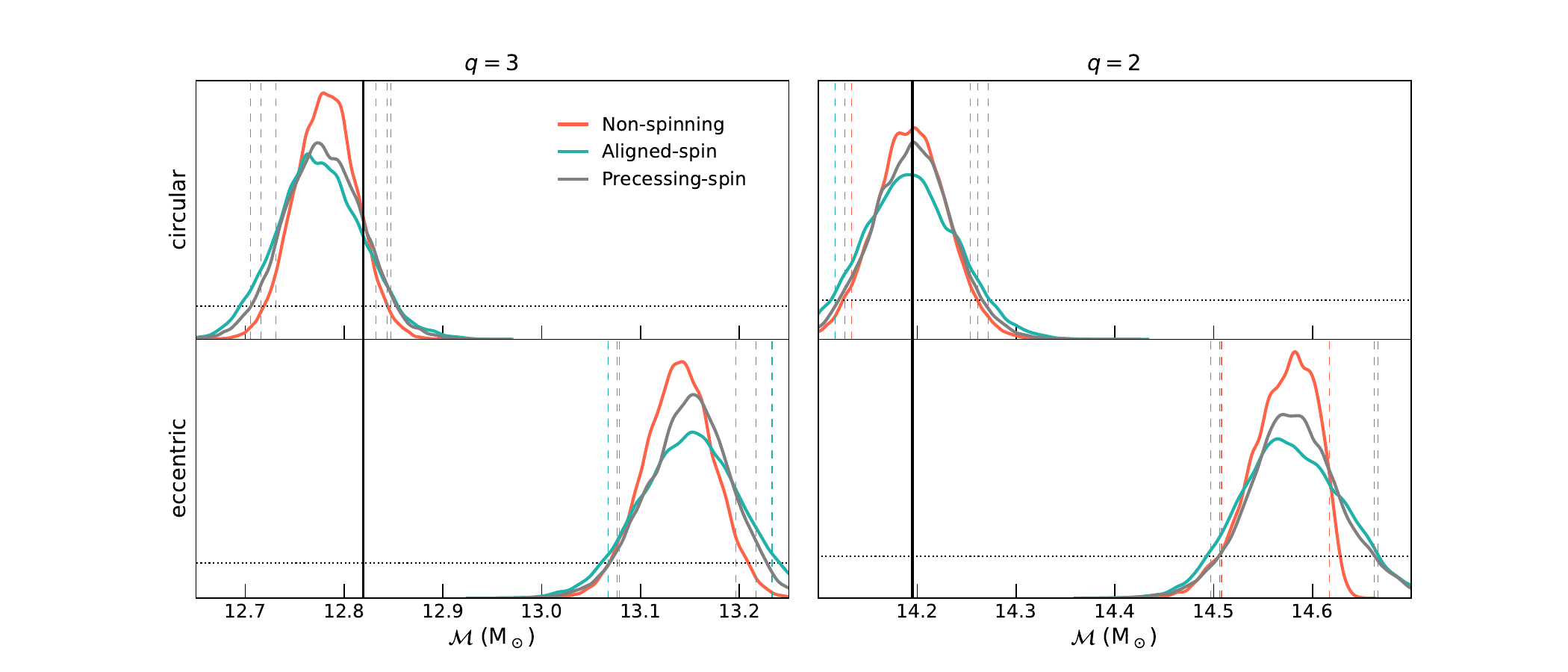}
    \includegraphics[trim=20 10 30 30, clip, width=0.275\linewidth]{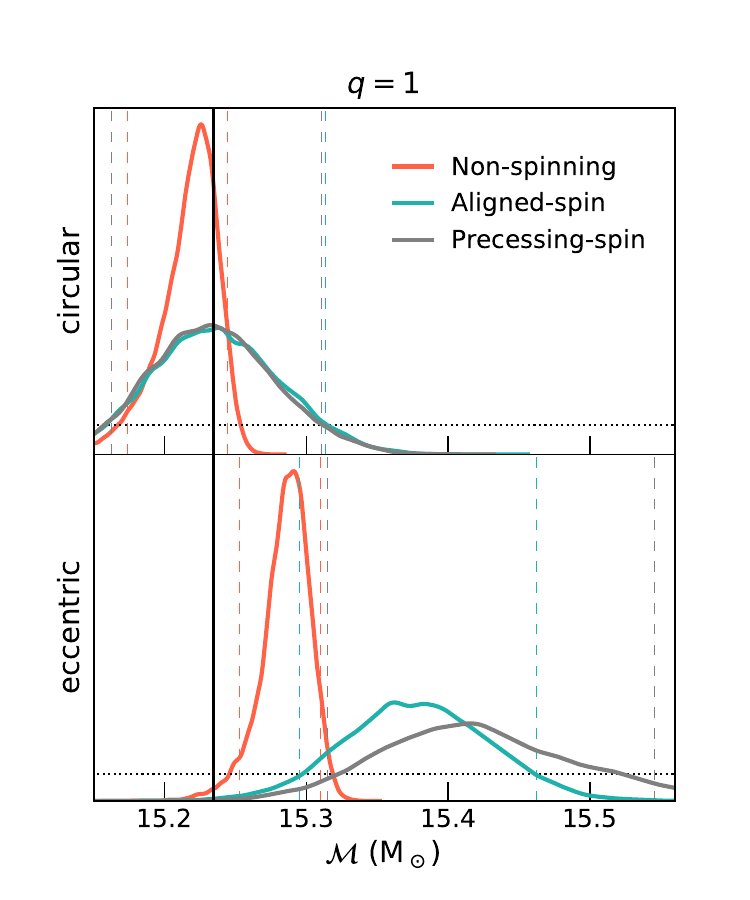}
    \caption{Chirp mass posteriors for injections with mass ratios ($q=1,2,3$). The rows indicate the nature of the injections, with the top panel showing results for the quasi-circular injection, and the bottom panel showing results for the eccentric injection ($e_{20} \sim 0.1$). The colours correspond to different spin settings used during recovery. Recovery is performed using quasi-circular waveforms in all cases: \textsc{IMRPhenomXAS} is used for the non-spinning (red) and aligned spin (green) recoveries, and \textsc{IMRPhenomXP} is used for recovery allowing precessing spins (grey). The dashed vertical coloured lines of the same colours denote the $90\%$ credible interval of the corresponding recoveries, the solid black line shows the injected value of $\mathcal{M}$, and the dotted black curve indicates the prior which is same for all recoveries. The injected value is recovered within the $90\%$ credible interval for the quasi-circular injections, while it is not recovered for the eccentric injections. The slight shift of posteriors for the quasi-circular injection in the $q=3$ case may be attributed to systematic differences between the waveform models used for injection and recovery. The matched filter SNRs for $q=1, 2$, and $3$ are $41, 38$, and $33$ respectively.}
    \label{fig:hist-ns-as-ps}
\end{figure*}

\begin{figure*}[ht!]
    \centering
    \includegraphics[width=0.32\linewidth]{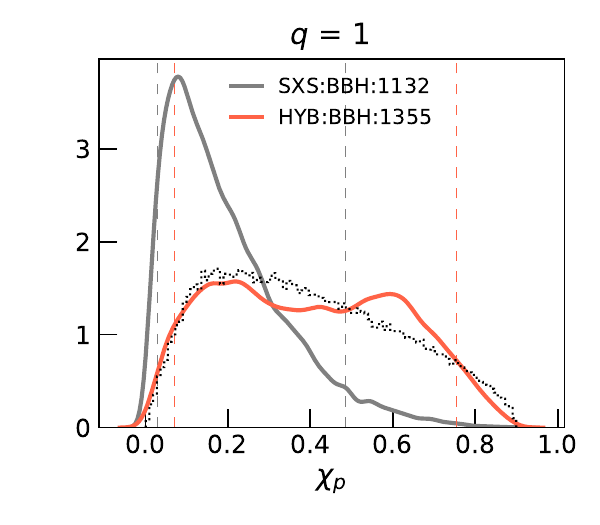}
    \includegraphics[width=0.32\linewidth]{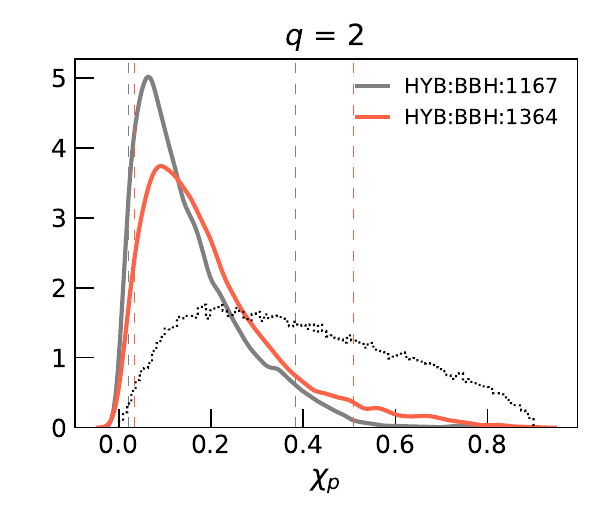}
    \includegraphics[width=0.32\linewidth]{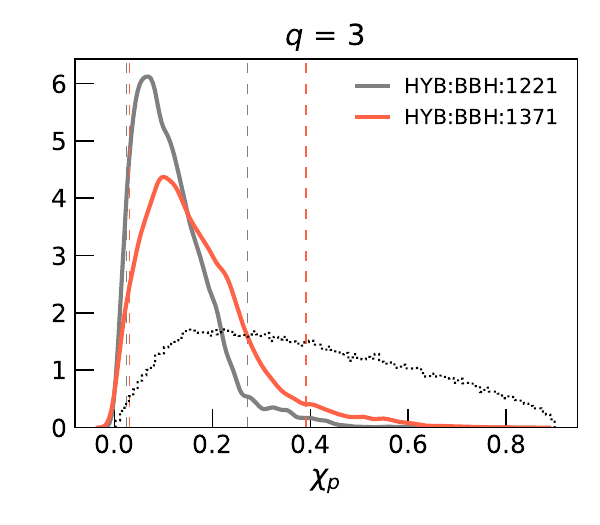}
    \caption{$\chi_p$ posteriors for non-spinning injections with mass ratios ($q=1,2,3$). The colours correspond to quasi-circular (grey) and eccentric (red) injections. Recovery is performed using \textsc{IMRPhenomXP}. The dashed vertical coloured lines of the same colours denote the $90\%$ credible interval of the corresponding injections, and the black dotted curve shows the prior which is same for both the injections.}
    \label{fig:hist_chip_ns_inj}
\end{figure*}

\begin{figure}[t!]
    \centering
    \includegraphics[trim=10 0 0 0, clip, width=\linewidth]{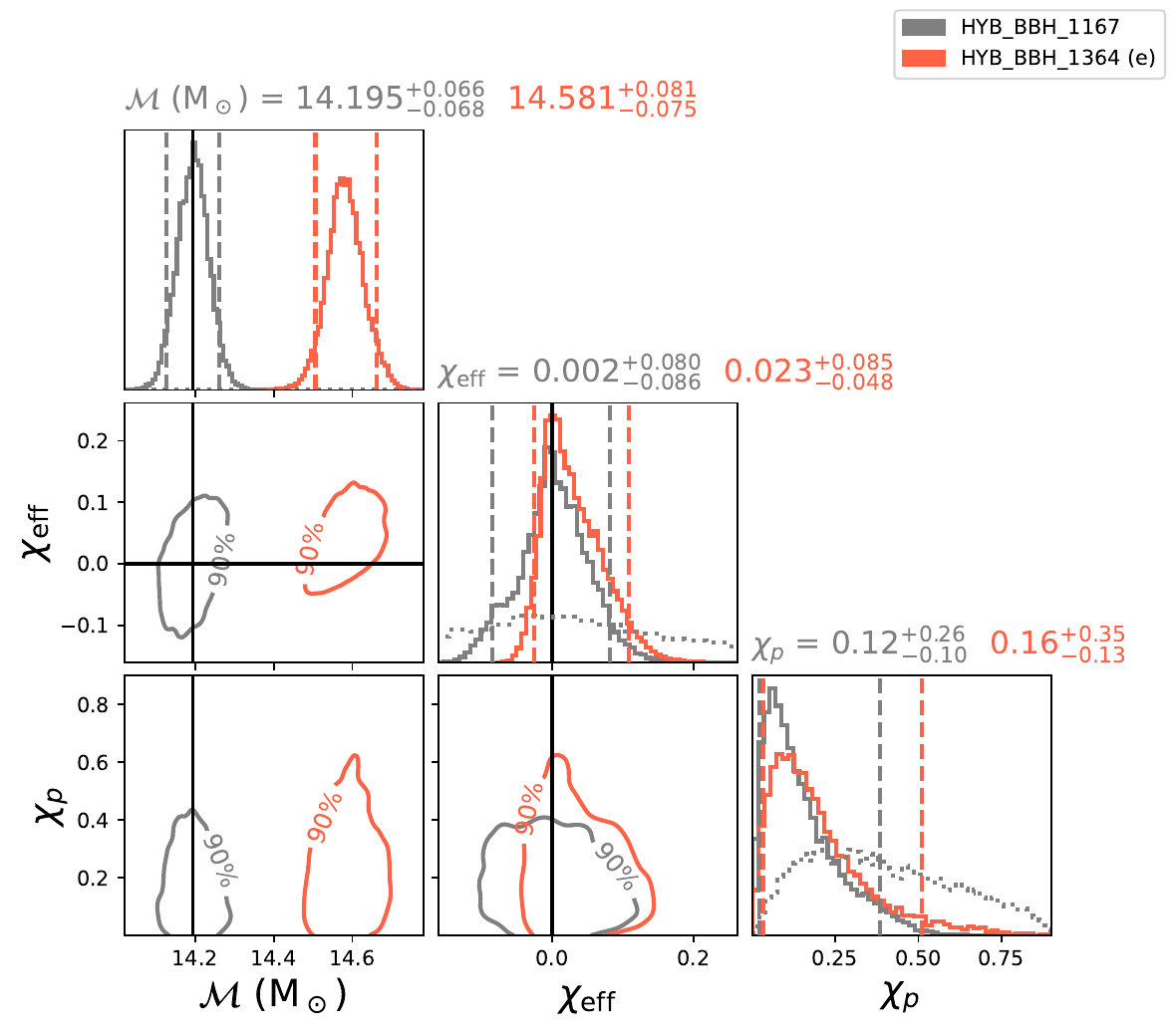}
    \caption{The corner plot for the $q=2$ injection, showing posteriors on the  chirp mass ($\mathcal{M}$), effective spin ($\chi_\text{eff}$), and spin precession parameter ($\chi_\text{p}$), for the precessing-spin recovery (performed using \textsc{IMRPhenomXP}) of both quasi-circular and eccentric injections. We show on the same plot results for the quasi-circular injection (grey), results for the eccentric injection (red), and the injection values black lines. The histograms shown on the diagonal of the plot are 1D marginalized posteriors for the respective parameters with vertical dashed lines denoting $90\%$ credible intervals. The dotted curves in the 1D plots show the priors used,\footnote{The prior height for $\mathcal{M}$ is too little compared to the posterior hence it is not visible in this plot. Hence, we show the $\mathcal{M}$ prior in Fig. \ref{fig:hist-ns-as-ps} instead.} which are same for the recovery of both quasi-circular and eccentric injections.
    }
    \label{fig:corner-ps-q-2}
\end{figure}

We perform zero-noise injections using non-spinning, quasi-circular as well as quasi-elliptical GW waveforms for BBH mergers of total mass of $M=35$~M$_\odot$ with mass ratios $q=(1, 2, 3$). These injections include the dominant modes ($\ell=2, |m|=2$) of the eccentric and quasi-circular IMR hybrids constructed in Ref.~\citep{Chattaraj:2022tay}, in addition to a quasi-circular SXS simulation (\textsc{SXS:BBH:1132}). Details of the simulations used in this study, including their eccentricity at the reference frequency of $20$~Hz, are shown in Table \ref{table:hybrids}. To calculate the eccentricity at $20$~Hz GW frequency, we have used the reference value ($e_0$) from Table I of Ref.~\cite{Chattaraj:2022tay} which they have quoted for a dimensionless frequency of $x=0.045$. Using the following relation we compute the $22$ mode frequency corresponding to $x=0.045$ and total mass $35$ $M_\odot$:
\begin{equation}
    f_{22} = \frac{x^{3/2}}{\pi M} = 17.62~\text{Hz}
\end{equation} 
where $M$ is the total mass taken in natural units (seconds). Now that we have $e_0$ at $f_{22}$, we evolve it using Eq.~(4.17a) of Ref.~\cite{Moore:2016qxz} (eccentricity evolution for orbit averaged frequency) to get eccentricity value at $20$~Hz ($e_{20}$ shown in Table \ref{table:hybrids}). Note that this is also the starting frequency for likelihood calculation.

\subsubsection{Quasi-circular, IMR recovery}

\begin{figure*}[t!]
    \centering
    \includegraphics[trim=10 0 10 10, clip, width=0.48\linewidth]{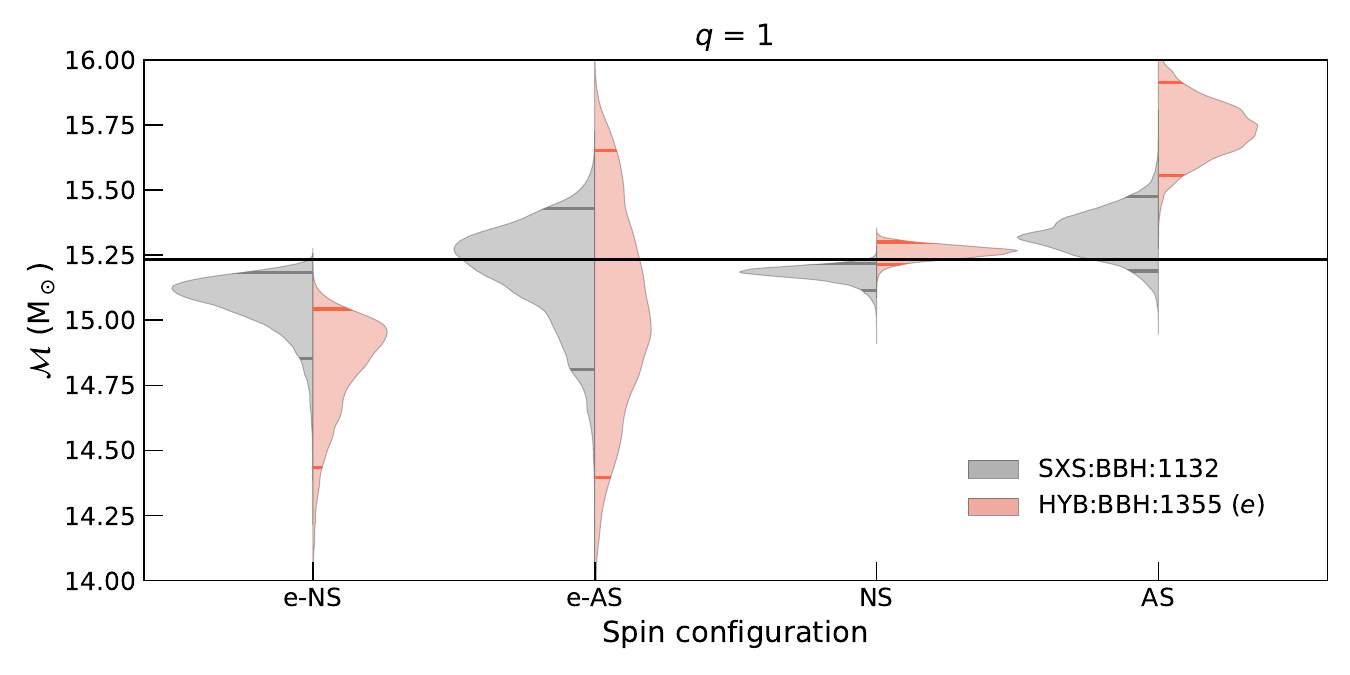}
    \includegraphics[trim=10 0 10 10, clip, width=0.48\linewidth]{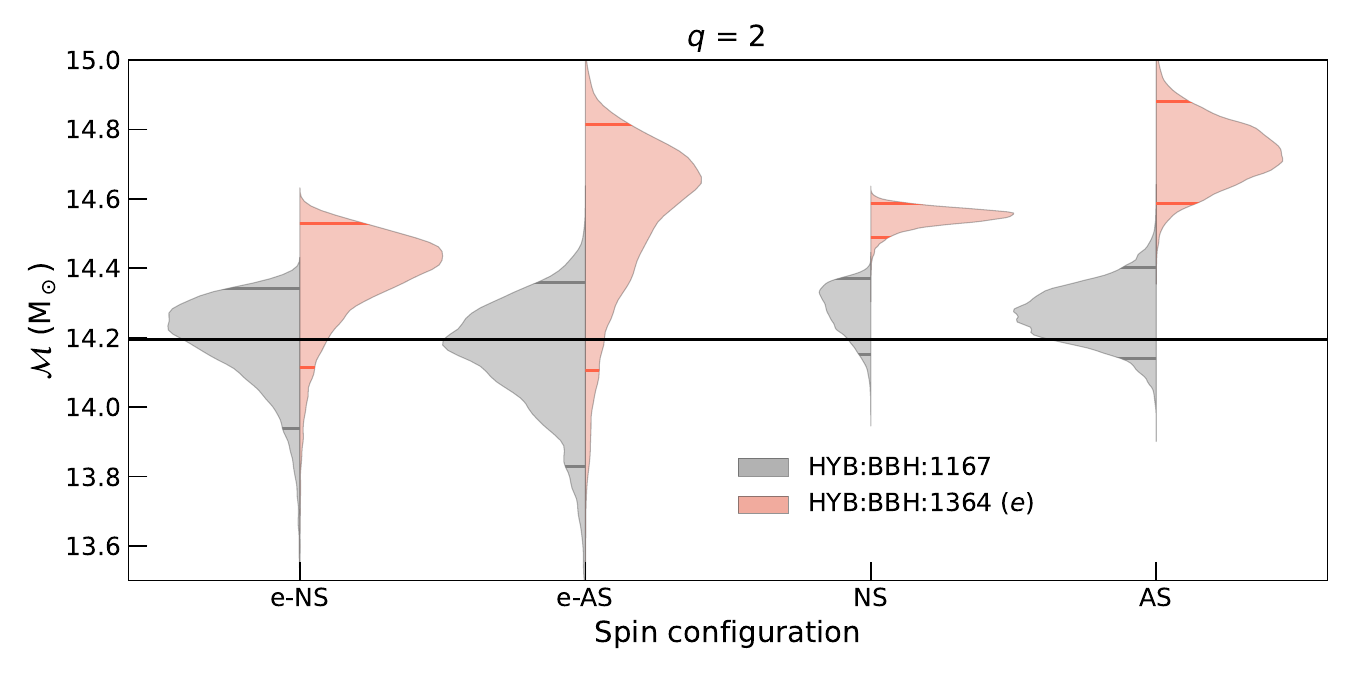}
    \includegraphics[trim=10 0 10 10, clip, width=0.48\linewidth]{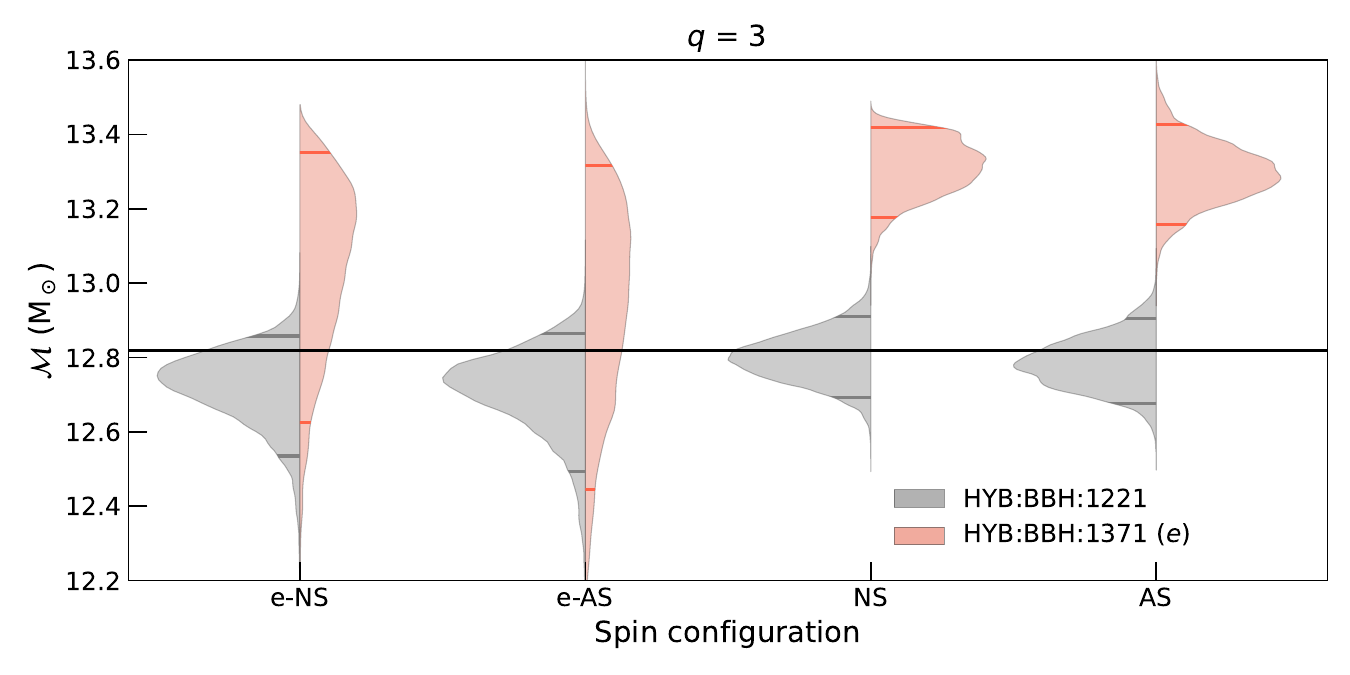}
    \caption{We show the recovery with \textsc{TaylorF2Ecc} for eccentric and quasi-circular injections in the form of violin plots, for mass ratios $q=1,2,3$. The colours used distinguish the injection hybrids: red shows eccentric ($e_{20} \sim 0.1$) injections while grey shows quasi-circular injections. The horizontal axis denotes the spin configuration in the recovery of posteriors. e-NS, e-AS, NS, and AS correspond to eccentric non-spinning, eccentric aligned spin, quasi-circular non-spinning, and quasi-circular aligned spin recoveries, respectively. The vertical axis corresponds to chirp mass values $\mathcal{M}$. The black horizontal line indicates the injection value and coloured lines inside the shaded posteriors indicate the $90\%$ credible interval. The matched filter SNRs for $q=1, 2$ and $q=3$ are $33, 31$ and $28$ respectively.}
    \label{fig:violin-ns-inj}
\end{figure*}

\begin{figure}
    \centering
    \includegraphics[width=\linewidth]{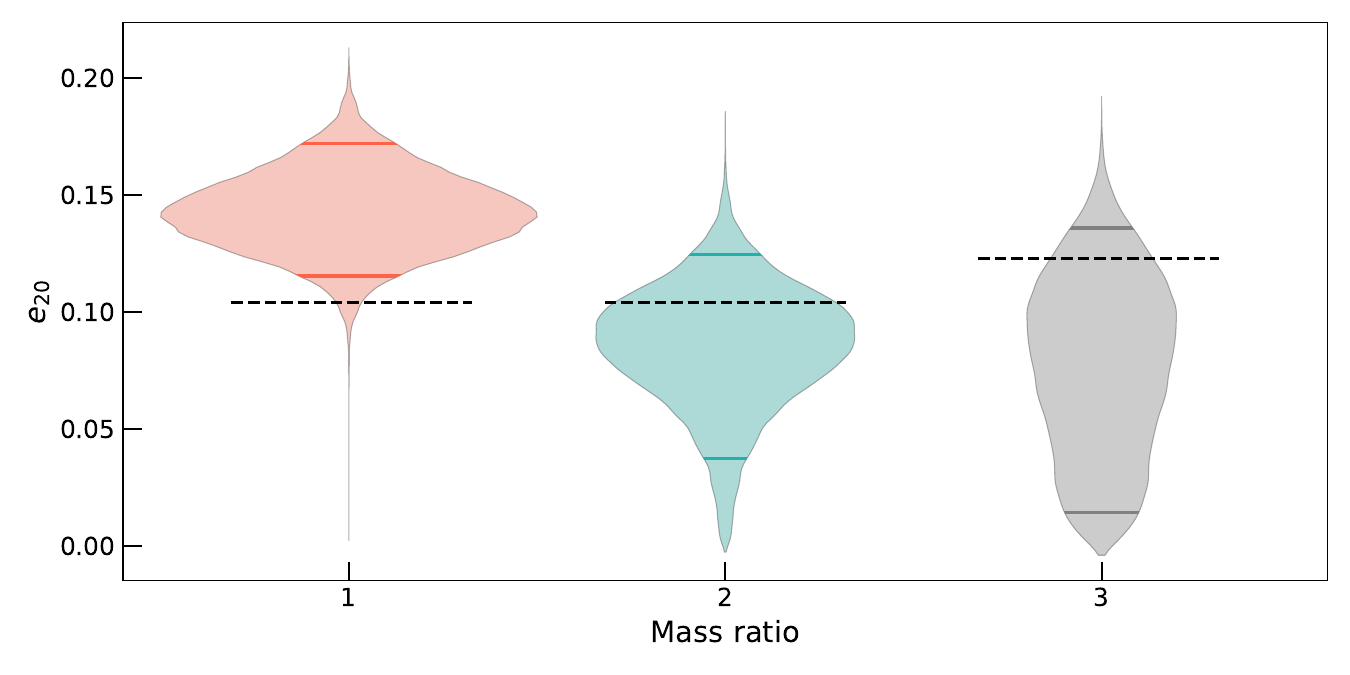}
    \caption{Eccentricity posteriors for $q$=(1, 2, 3) when the injections are non-spinning and eccentric, and the recovery is with \textsc{TaylorF2Ecc} in eccentric, non-spinning configuration. The coloured lines inside the shaded posteriors indicate 90\% credible interval whereas the black dashed lines denote the injected eccentricity values.}
    \label{fig:ecc_ns_inj_q_1_2_3}
\end{figure}

Here we explore the bias introduced in the source parameters recovered via parameter estimation (PE) of GW events when eccentricity is ignored, i.e. we inject an eccentric signal but do not use eccentric waveforms for recovery. For this exercise, we use the quasi-circular IMR phenomenological waveform models (\textsc{IMRPhenomXAS}\citep{Pratten:2020fqn} and \textsc{IMRPhenomXP}\citep{Pratten:2020ceb}). 
\\
\\
In Fig.~\ref{fig:hist-ns-as-ps}, we plot the posterior probability distributions on chirp mass $\mathcal{M}$ for non-spinning quasi-circular and eccentric injections recovered using IMR quasi-circular waveform models. The injection is recovered using three spin setting configurations: non-spinning (NS), in which we restrict all spins to $0$; aligned-spin (AS), in which we restrict spin-tilt angles to $0$ and allow spin magnitudes to range between $0$ and $0.99$; and precessing-spin (PS), in which we allow all spin parameters to vary. By using different spin configurations, we investigate whether spurious measurements of spin occur for non-spinning eccentric injections. The spin precession parameter $\chi_p$ is plotted in Fig.~\ref{fig:hist_chip_ns_inj} for PS recovery. In addition to studying biases in spin parameters due to the presence of eccentricity, we also compare the effect of spin settings on the recovery of chirp mass in the presence of eccentricity, since eccentricity and chirp mass are known to be correlated parameters (see for instance \citet{Favata:2021vhw}). We use the waveform \textsc{IMRPhenomXAS} for non-spinning and aligned-spin recoveries, and \textsc{IMRPhenomXP} for the precessing-spin recovery. We display the corresponding corner plots in Figs.~\ref{fig:corner-ps-q-2} and \ref{fig:corner-ps-q-1-3} in Appendix \ref{appendix:q_1} for chirp mass ($\mathcal{M}$), effective spin parameter ($\chi_\text{eff}$), and spin precession parameter ($\chi_p$). Looking at figures \ref{fig:hist-ns-as-ps}, \ref{fig:hist_chip_ns_inj}, \ref{fig:corner-ps-q-2}, and \ref{fig:corner-ps-q-1-3} in Appendix \ref{appendix:q_1}, we make the following observations (also seen in previous studies such as \cite{Ramos-Buades:2019uvh, Moore:2019vjj}):

\begin{itemize}
    \item In Fig.~\ref{fig:hist-ns-as-ps}, it can be noted that for all the values of mass ratio that we consider, the recovery of eccentric injections with quasi-circular waveform models results in a significant bias of the chirp mass posterior, such that the injected value falls outside the $90\%$ credible interval. This leads us to conclude that when eccentric signals are analysed using quasi-circular waveform models, these models are unable to capture the true parameters of the eccentric signals.
    
    \item In the same figure, we also observe that the spin settings (non-spinning, aligned-spin, or precessing-spin) chosen for recovery do not affect the magnitude of shift in the chirp mass posterior for mass ratios 2 and 3, or in other words; the bias in the recovered $\mathcal{M}$ is same regardless of assumptions about spin magnitude and spin tilt. This leads us to conclude that including spins in parameter estimation with quasi-circular waveform models has negligible impact on the recovery of the chirp mass posterior when the injection is non-spinning and eccentric in nature.
    
    \item For $q=1$, the shift in the chirp mass posteriors for different spin configurations, seen in the right panel of Fig.~\ref{fig:hist-ns-as-ps}, can partly be explained due to the prior railing of mass ratio ($q$) leading to almost the entire posterior volume lying outside the injected value. This can lead to prior railing in component masses and other correlated parameters. This is discussed in detail in appendix \ref{appendix:q_1}.
        
    \item The effective spin ($\chi_\text{eff}$) posteriors, seen in Figs.~\ref{fig:corner-ps-q-2} and \ref{fig:corner-ps-q-1-3}, are largely consistent with zero for both quasi-circular and eccentric injections, for $q=2$ and $q=3$ cases. The slight deviation of $\chi_\text{eff}$ from $0$ for $q=1$ case can be explained by looking at the correlation between chirp mass and $\chi_\text{eff}$ (see Fig.~\ref{fig:corner-ps-q-1-3} in appendix \ref{appendix:q_1}). We also note in Fig.~\ref{fig:corner-ps-q-2}, looking at the 2D posteriors, that chirp mass shows correlation with $\chi_\text{eff}$ both for quasi-circular and eccentric injections. On the other hand, there doesn't seem to be direct and visible correlation of $\chi_\text{p}$ with $\mathcal{M}$ and $\chi_\text{eff}$.
    
    \item In Fig.~\ref{fig:hist_chip_ns_inj}, it can be seen that the posteriors for $\chi_\text{p}$ peak toward 0 for the $q=2$ and $q=3$ cases. Since $\chi_\text{p}$ is a measure of the misalignment of spins in a binary system, these posteriors peaking toward zero indicate little to no evidence of spin-precession in the signal. For $q=1$, the $\chi_p$ posterior for eccentric injection is uninformative. Both the uninformative posterior and posteriors that peak toward $0$ support the conclusion that \textbf{eccentricity is not confused for spin-induced precession in long-duration signals from low-mass BBH}.
\end{itemize}

For the $q=3$ case, the slight deviation of posteriors from the injected value for the quasi-circular injection (top left panel of Fig.~\ref{fig:hist-ns-as-ps}) is most likely due to systematic differences between the injection and the recovery waveform. Even at a reasonably modest eccentricity of $e_{20}\sim0.1$, the chirp mass posterior is shifted enough that the injected value is not recovered within $90\%$ confidence. Further, for a non-spinning eccentric system with moderate total mass ($M=35 M_\odot$), the presence of eccentricity in the signal is not mimicked by a spin-precessing quasi-circular waveform. This is consistent with the findings of \citet{Romero-Shaw:2022fbf}, who find that eccentricity and spin-precession may be distinguished in signals with long inspirals coming from low-mass BBH due to the signal duration exceeding the timescale upon which modulations induced by eccentricity differ significantly from those induced by spin-precession. The fact that the spin posteriors are similar for both eccentric and quasi-circular injections also implies that a lack of spin can be confidently identified in low-mass systems regardless of their eccentricity.

\subsubsection{Eccentric, inspiral-only recovery}

\begin{figure*}[t!]
    \centering
    \includegraphics[trim=10 0 10 10, clip, width=0.48\linewidth]{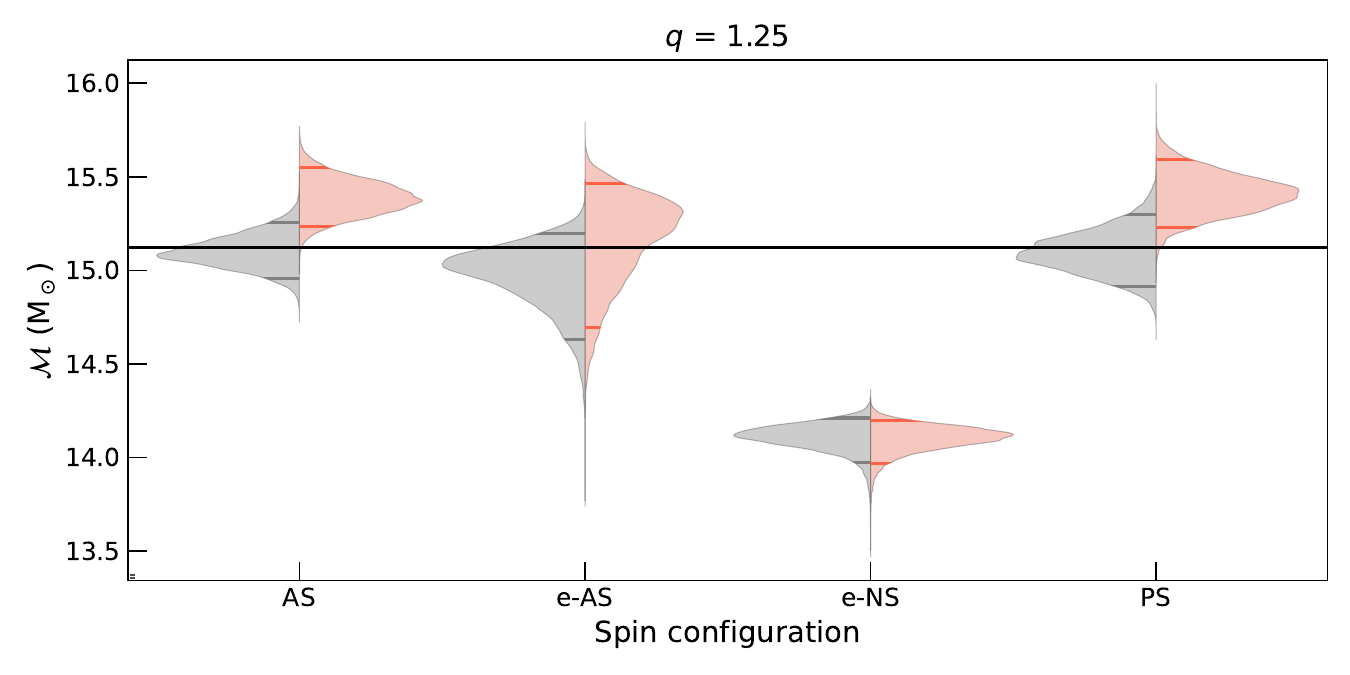}
    \includegraphics[trim=10 0 10 10, clip, width=0.48\linewidth]{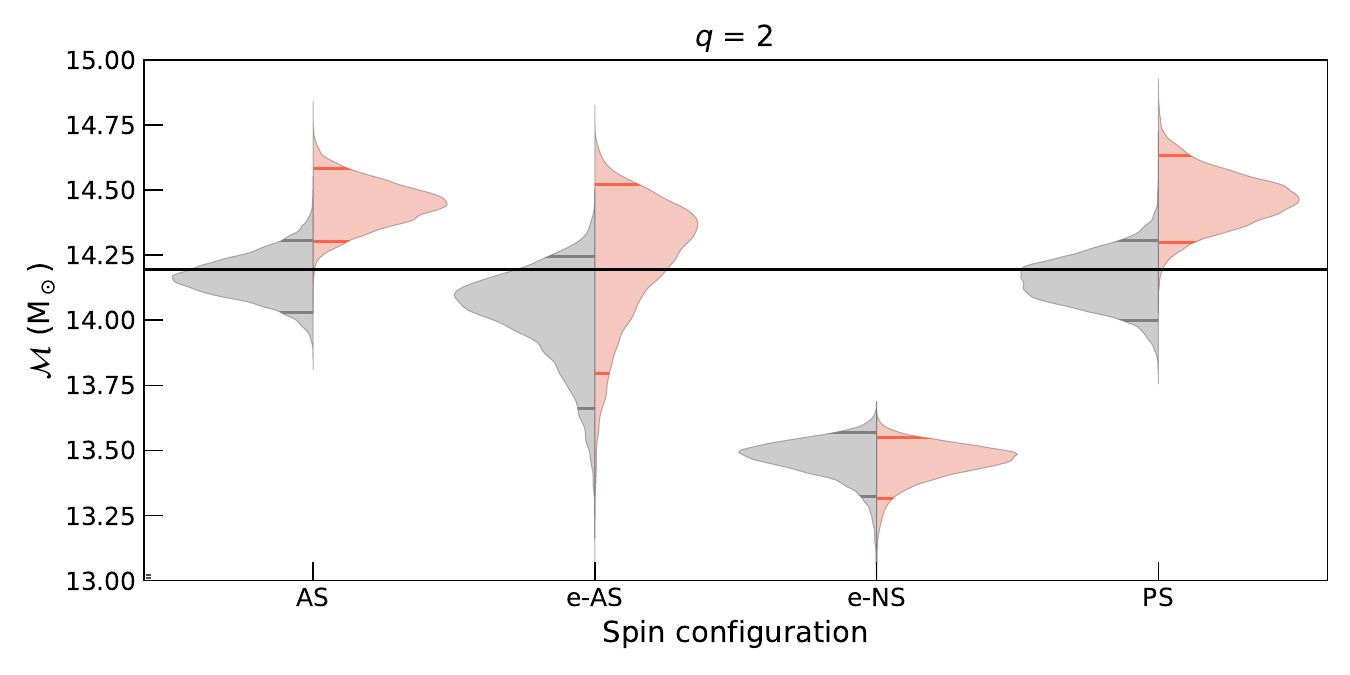}
    \includegraphics[trim=10 0 10 10, clip, width=0.48\linewidth]{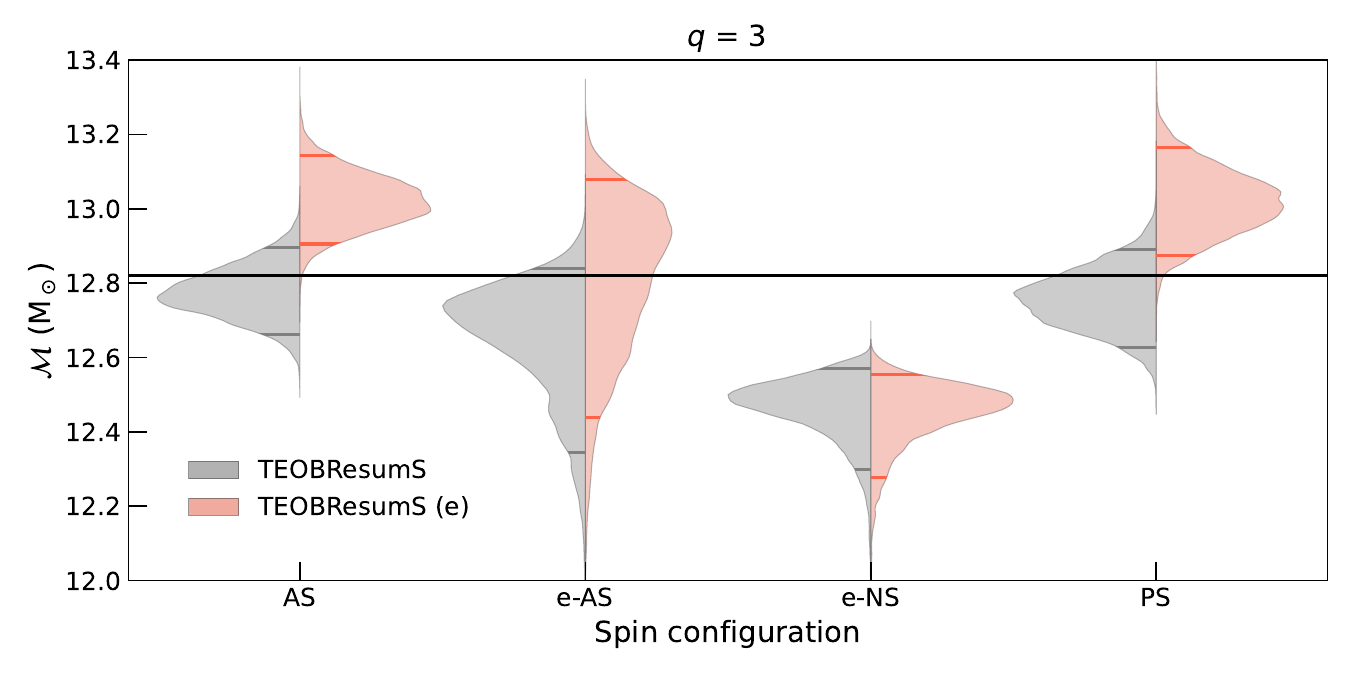}
    \caption{Eccentric ($e_{20}=0.1$) (red) and quasi-circular (grey) aligned-spin injections generated with eccentric waveform model \textsc{TEOBResumS}, recovered with \textsc{TaylorF2Ecc} in quasi-circular aligned-spin (AS), eccentric aligned-spin (e-AS), and eccentric non-spinning (e-NS) configurations, and with quasi-circular waveform model \textsc{IMRPhenomXP} including precessing spin (PS). The aligned-spin eccentric injection is only correctly recovered using the e-AS configuration in the recovery waveform. The matched filter SNRs for $q=1.25$, $q=2$, and $q=3$ are $35$, $33$, and $30$ respectively.}
    \label{fig:violin-as-inj}
\end{figure*}

We analyze the same injections as in the previous section with an eccentric inspiral-only waveform model \textsc{TaylorF2Ecc}. We present the results in Figs.~\ref{fig:violin-ns-inj} and \ref{fig:ecc_ns_inj_q_1_2_3}. We also show results obtained when the recovery is performed under the constraint that $e_{20}=0$ with the same waveform, in order to account for any biases arising due to systematic differences between waveform model families and/or the lack of merger and ringdown in \textsc{TaylorF2Ecc}. Since \textsc{TaylorF2Ecc} is an inspiral-only model, we have truncated the likelihood integration in the recovery using the quasi-circular waveform model also to the same frequency as the eccentric recovery for a fair comparison and to get comparable SNRs. This frequency has been chosen to be $110$~Hz, close to the ISCO frequency for a $35$~M$_\odot$ system.\\
\\
In the case of mass ratios $q=2,3$, for eccentric injections (plotted in red in the figures), a quasi-circular recovery excludes the injection value of chirp mass from the $90\%$ credible interval when the recovery waveform has arbitrary spin constraints, whereas when eccentricity is included in the recovery waveform model and is sampled over, the injected value is recovered within the $90\%$ credible interval. We show the 1D posteriors for eccentricity in Fig.~\ref{fig:ecc_ns_inj_q_1_2_3}, and the 2D contours of $e_{20}$ with $\mathcal{M}$ and $q$ in Fig.~\ref{fig:corner-ns-ecc} of Appendix \ref{appendix:ecc_corner} which highlights the correlations between these parameters. The $\log$-Bayes factors between eccentric and quasi-circular recoveries performed with \textsc{TaylorF2Ecc} are close to $0$ for $q=2$ and $q=3$, but for $q=1$ it is $\sim11$, and thus favours recovery with an eccentric template when the injected waveform is eccentric. 
However, for $q=1$ (top left panel of Fig.~\ref{fig:violin-ns-inj}) even the quasi-circular recovery (NS) for quasi-circular injection (grey) is not recovered within $90\%$ confidence. This may be partially caused by waveform systematics between the injected and recovered waveforms. Additionally, as shown in Fig.~\ref{fig:ecc_ns_inj_q_1_2_3}, the injected $e_{20}$ is not recovered within $90\%$ confidence for the eccentric injection. As well as waveform systematics, this is likely because the injected $q$ is at the lower boundary of the prior, so the entire $q$ posterior spans higher $q$ than injected, and (as shown in, for example, Fig.~\ref{fig:corner-ns-ecc}) higher $q$ correlates with higher $e_{20}$. To eliminate possible biases due to this prior effect, we use $q=1.25$ instead of $q=1$ in the following section.

\subsection{Aligned-spin, eccentric injections}
\label{subsec:align-spin-inj}

\begin{figure*}[ht!]
    \centering
    \includegraphics[width=0.32\linewidth]{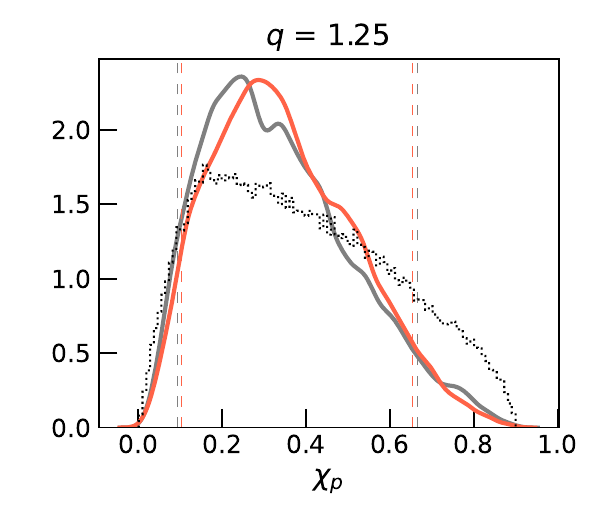}
    \includegraphics[width=0.32\linewidth]{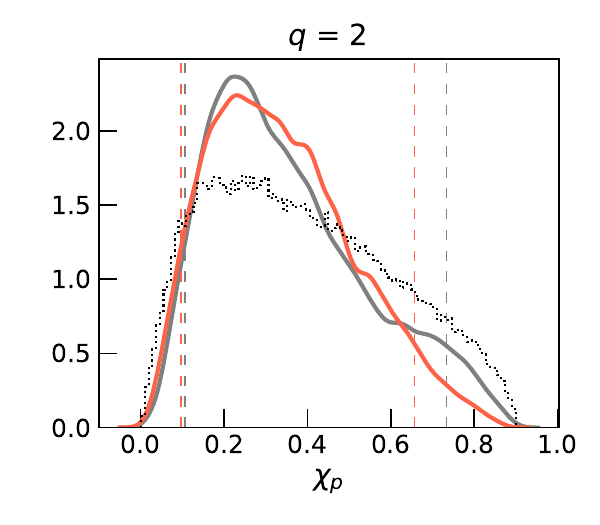}
    \includegraphics[width=0.32\linewidth]{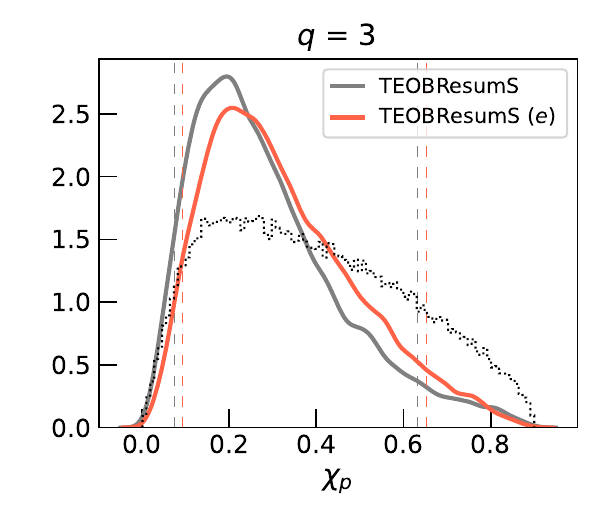}
    \caption{$\chi_p$ posteriors for aligned spin injections with mass ratios ($q=1.25,2,3$). The colours correspond to quasi-circular (grey) and eccentric (red) injections. Recovery is performed using \textsc{IMRPhenomXP} truncated at 110~Hz for PE. The dashed vertical coloured lines of the same colours denote the $90\%$ credible interval of the corresponding injections, and the black dotted curve shows the prior which is same for both the injections.}
    \label{fig:hist_chip_as_inj}
\end{figure*}

\begin{figure}[t!]
    \centering
    \includegraphics[width=\linewidth]{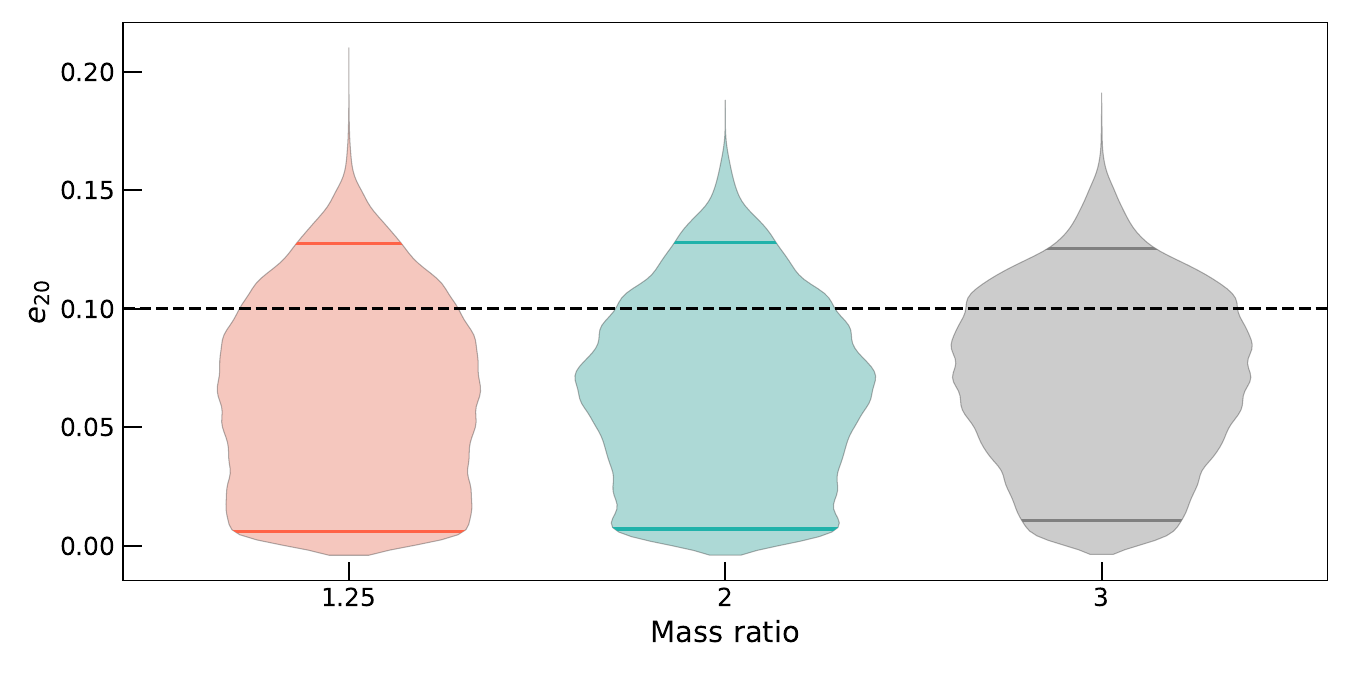}
    \caption{Eccentricity posteriors for $q$=(1.25, 2, 3) when the injections are aligned-spin and eccentric, and the recovery is with \textsc{TaylorF2Ecc} in eccentric, aligned-spin configuration. The coloured lines inside the shaded posteriors indicate 90\% credible interval whereas the black dashed lines denotes the injected eccentricity value.}
    \label{fig:ecc_as_inj_q_1.25_2_3}
\end{figure}

We inject an aligned-spin eccentric signal using the waveform model \textsc{TEOBResumS} \cite{Nagar:2018zoe} and recover in four configurations: quasi-circular aligned spin (AS), eccentric aligned spin (e-AS), eccentric non-spinning (e-NS), and quasi-circular precessing spin (PS). For the first three cases (AS, e-AS, and e-NS) we have used the waveform model \textsc{TaylorF2Ecc} for recovery, whereas for the last case (PS), we have used \textsc{IMRPhenomXP}. Since \textsc{TaylorF2Ecc} is an inspiral-only waveform, we truncate the likelihood calculation at $110$~Hz in accordance with the choice of total mass as described above. As above, the total mass is $35$ M$_\odot$, and here we choose to inject signals with mass ratios $q=1.25, 2, 3$. The injected spin magnitudes are $\chi_\text{1z} = \chi_\text{2z} = 0.3 = \chi_\text{eff}$. 
The eccentric injections have $e_{20}=0.1$, consistent with injections in earlier sections, with the eccentricity defined at the orbit-averaged frequency of $20$ Hz \cite{TEOBResumS:bitbucket}. The chirp mass posteriors for the above are plotted in the form of violin plots in Fig.~\ref{fig:violin-as-inj}. For arbitrary spin settings, the quasi-circular waveform models are not able to correctly recover the chirp mass of the eccentric injection, whereas the eccentric waveform model with aligned spins does correctly recover the injected value. 
This once again indicates that a quasi-circular precessing waveform model can cause biases in the recovered value of $\mathcal{M}$ if the signal is truly eccentric. Further, looking at Figs.~\ref{fig:hist_chip_as_inj} and \ref{fig:corner-ps-q-2-3-as-inj}, we note that the $\chi_p$ posteriors peak at the same value for both eccentric and quasi-circular injections. This again suggests that an aligned spin eccentric signal when recovered with quasi-circular precessing model does not mimic a spin-precessing signal any more than a quasi-circular aligned-spin injection.
\\
\\
We also analyze the same injection with the eccentric waveform model with zero spins (shown as e-NS in the figures). In this case, the posteriors for both the quasi-circular and eccentric injections are biased toward lower masses than injected. A positively aligned-spin system has more cycles compared to its non-spinning counterpart, which is also the case for lower-mass systems. Thus, when an aligned-spin signal is recovered using a non-spinning waveform model, it is naturally biased toward lower masses\footnote{We note here that this bias in the chirp mass also causes bias in the eccentricity posterior, resulting in a higher value of eccentricity due to negative correlation with chirp mass. This can result in posteriors peaking at non-zero values of eccentricities when an aligned spin quasi-circular signal is analyzed with non-spinning eccentric model.}. Hence, a system which is eccentric and spinning (aligned) may be recovered with a positive bias in chirp mass when eccentricity is ignored, and a negative bias when the spins are ignored. Again, we provide the eccentricity posteriors in Fig.~\ref{fig:ecc_as_inj_q_1.25_2_3} as well as in the form of corner plots along with $\mathcal{M}$, $q$, and $\chi_\text{eff}$ parameters in Fig.~\ref{fig:corner-as-ecc} of Appendix \ref{appendix:ecc_corner}. Looking at Fig.~\ref{fig:corner-as-ecc}, there is both a clear correlation between $e_{20}$ and $\mathcal{M}$, and a mild correlation between $\chi_\text{eff}$ and $e_{20}$, consistent with the findings of \citet{OShea:2021faf}. Also, as in Sec.~\ref{subsec:non-spin-inj}, we find that the Bayes factors ($\mathcal{B}_{E/C}$) for $q=1.25$, $2$, and $3$ are not high enough to indicate a clear preference for either the quasi-circular or eccentric waveform model for any injection.

%----------------------------------------------------------------------------------------------------------

\section{Conclusions} 
\label{sec:concl}
Measurable orbital eccentricity is a key indicator of BBH formation channel. However, catalogs of BBH detections, e.g. \citet{KAGRA:2021vkt}, typically neglect this parameter and study all GW candidates using only quasi-circular signal models. Additionally, matched-filter searches for GW signals typically rely on quasi-circular waveform templates. In this work, we explore both the detectability of the eccentric signals when eccentricity is neglected from matched-filtering searches, and the biases that result from performing parameter estimation on eccentric GW signals using quasi-circular waveform models under a variety of spin assumptions. 

We find that there is a loss in the fitting factor ($< 0.95$) for eccentricities higher than $0.01$ at $10$~Hz in conjunction with high values of mass ratio ($q>3$). Further, we calculate the signal recovery fraction (SRF) and find that there's a loss in SRF up to 6\% for the region in parameter space with $e_{10} > 0.01$ and mass ratio $q > 3$. While we restrict this calculation to the inspiral section of the signal, we argue that the loss in $FF$ would be similar for full IMR signals, since eccentricity is efficiently radiated away from an inspiralling system and so the binary should be close to circular before the merger and ringdown. 
The overall loss in the fraction of recovered signals depends on the fraction of events in the population that are high-eccentricity and high-mass ratio. 
These population characteristics, in turn, depend on the formation channels contributing to the population.
For example, we would detect a higher fraction of binaries that formed in globular clusters (GCs) than those that formed in active galactic nuclei (AGN), since the eccentricity and mass ratio distributions expected from binaries forming in GCs are less extreme than those expected from AGN \citep[e.g.,][]{Zevin:2019:EccentricGCs, Tagawa:2021:MassGap, Tagawa:2021:AGN, Yang:2019cbr}.
Therefore, with a severity depending on the balance between the formation channels contributing to the observed population, missing eccentric binaries in searches can lead to errors in the inferred merger rate and underlying population characteristics. We note that the percentage of the population recovered is likely to vary slightly if the second eccentric parameter is also varied in the injected population. However, we expect that the variation would be moderate when averaged over the full possible range of this parameter.\\
\\
Even if an eccentric signal is detected via a matched-template search based on quasi-circular waveform templates, the recovery of source parameters can be biased when the signal is analysed using a quasi-circular waveform model. We perform parameter estimation on non-spinning and aligned-spin eccentric injections, and recover them using various spin assumptions with quasi-circular and eccentric waveform models. We find that for $e_{20}\sim 0.1$, analyses with the quasi-circular waveform models are unable to recover the injected values of chirp mass within the $90\%$ credible interval. Further, we note that for the relatively low-mass BBH considered in this study, no spurious spin detections are made for non-spinning eccentric injections, and no spurious inferences of precession are made for any eccentric injections. This leads us to conclude that \textbf{for relatively low mass systems, spin-precession does not mimic eccentricity}. The spin parameter posteriors are similar for both quasi-circular and eccentric injections. 

As discussed in Sec.~\ref{sec:intro}, both eccentricity and spin-precession can indicate that a binary formed in a dynamical environment. Our results suggest that a non-spinning low-mass eccentric system, if analyzed using quasi-circular waveform models only, may be mistaken for a binary that formed in isolation since the quasi-circular waveform models do not enable measurements of eccentricity and the spin posteriors show no additional signatures of dynamical formation. This may lead to miscategorization of such systems as uninteresting or ``vanilla'' binaries. Moreover, if we are routinely biased to higher masses even for a small subset of signals that include the influence of binary eccentricity, the population distribution of mass will gradually be shifted higher. Eventually, this could lead to incorrect inferences about, for example, the location of the pair-instability mass gap and the fraction of the population comprised of hierarchical mergers \citep[using hierarchical inference methods such as, e.g.,][]{Mould:2022ccw}. While the shifts in the inferred chirp mass for the low-mass and moderate-eccentricity injections studied here are relatively minor, for higher eccentricities and higher masses the bias would likely be worse (see for instance Eq.~(1.1) in \cite{Favata:2021vhw}). 

We also observe that for the eccentricity values chosen here ($e_{20}\sim 0.1$), even an inspiral-only eccentric waveform with no second eccentricity parameter, e.g., mean anomaly, is able to recover the injected chirp mass within the $90\%$ confidence interval. Therefore, we conclude that for GW signals from relatively low-mass BBH, inspiral-only eccentric waveform models are adequate for identifying and quantifying orbital eccentricity.

\acknowledgments

We thank Patricia Schmidt for a critical reading of the manuscript and useful comments. We thank K.~G.~Arun, B.~S.~Sathyaprakash, Anuradha Gupta, Alexander H.~Nitz, and Rahul Dhurkunde for useful discussions. I.M.R-S. acknowledges support received from the Herchel Smith Postdoctoral Fellowship Fund.  C.~K.~M.~acknowledges the support of SERB's Core Research Grant No. CRG/2022/007959. Computations were performed on the \texttt{powehi} workstation in the Department of Physics, IIT Madras, ATLAS cluster at AEI Hannover, and CIT cluster provided by the LIGO Laboratory. The authors are grateful for computational resources provided by the LIGO Laboratory and supported by National Science Foundation Grants Np.~PHY-0757058 and No.~PHY-0823459. We used the following software packages: {\tt LALSuite}~\cite{lalsuite}, {\tt PyCBC}~\cite{alex_nitz_2020_4134752}, {\tt NumPy}~\cite{2020Natur.585..357H}, {\tt Matplotlib}~\cite{2007CSE.....9...90H}, {\tt Seaborn}~\cite{Waskom2021}, {\tt jupyter}~\cite{soton403913}, {\tt dynesty}~\cite{2020MNRAS.493.3132S}, {\tt corner}~\cite{corner}.
This document has LIGO preprint number {\tt LIGO-P2300326}.

\appendix

\section{Priors used for parameter estimation}
\label{appendix:priors}

The priors on various parameters used for non-spinning, aligned-spin, and precessing-spin analyses are listed in Table \ref{table:priors}.

\begin{table}
\def\arraystretch{1.4}
\begin{tabular}{|c|c|c|}
\hline
\textbf{Parameter} & \textbf{Prior} & \textbf{Range} \\ \hline
$\mathcal{M}$ & \begin{tabular}[c]{@{}c@{}}Uniform in \\ component masses \end{tabular} & $5 \text{ - } 50~M_\odot$ \\ \hline
$q$ & \begin{tabular}[c]{@{}c@{}}Uniform in \\ component masses \end{tabular} & $1 \text{ - } 5$ \\ \hline
$d_L$ & Uniform radius & $100 \text{ - } 3000$ Mpc \\ \hline
$\iota$ & Uniform sine & $0 \text{ - } \pi$ \\ \hline
$t_c$ & Uniform & $t_\text{gps}+(-0.1 \text{ - } 0.1)$~s \\ \hline
$\phi_c$ & Uniform & $0 \text{ - } 2\pi$ \\ \hline
$\chi_{i\text{z}}$\footnote{\label{note:index}where $i=[1,2]$ refers to the binary components}\textsuperscript{, }\footnote{only used for aligned-spin recovery} & Uniform & $0 \text{ - } 0.9$ \\ \hline
$a_1$, $a_2$\footnote{\label{note:prec-spin}only used for precessing spin recovery} & Uniform & $0 \text{ - } 0.9$ \\ \hline
$(S_i^\Theta + S_i^\Phi)$\textsuperscript{\ref{note:index}, \ref{note:prec-spin}} & Uniform solid angle & \begin{tabular}[c]{@{}c@{}} $\Theta \in (0,\pi)$, \\ $\Phi \in (0,2\pi)$ \end{tabular} \\ \hline
$(\alpha + \delta)$ & Uniform sky & \begin{tabular}[c]{@{}c@{}} $\delta \in (\pi/2,-\pi/2)$, \\ $\alpha \in (0,2\pi)$ \end{tabular} \\ \hline
$e$\footnote{only used for eccentric recovery} & Uniform & $0 \text{ - } 0.4$ \\ \hline
\end{tabular}
\caption{Priors for parameters in various quasi-circular and eccentric recoveries.}
\label{table:priors}
\end{table}

\begin{figure*}
    \centering
    \includegraphics[trim=10 0 0 0, clip, width=0.49\linewidth]{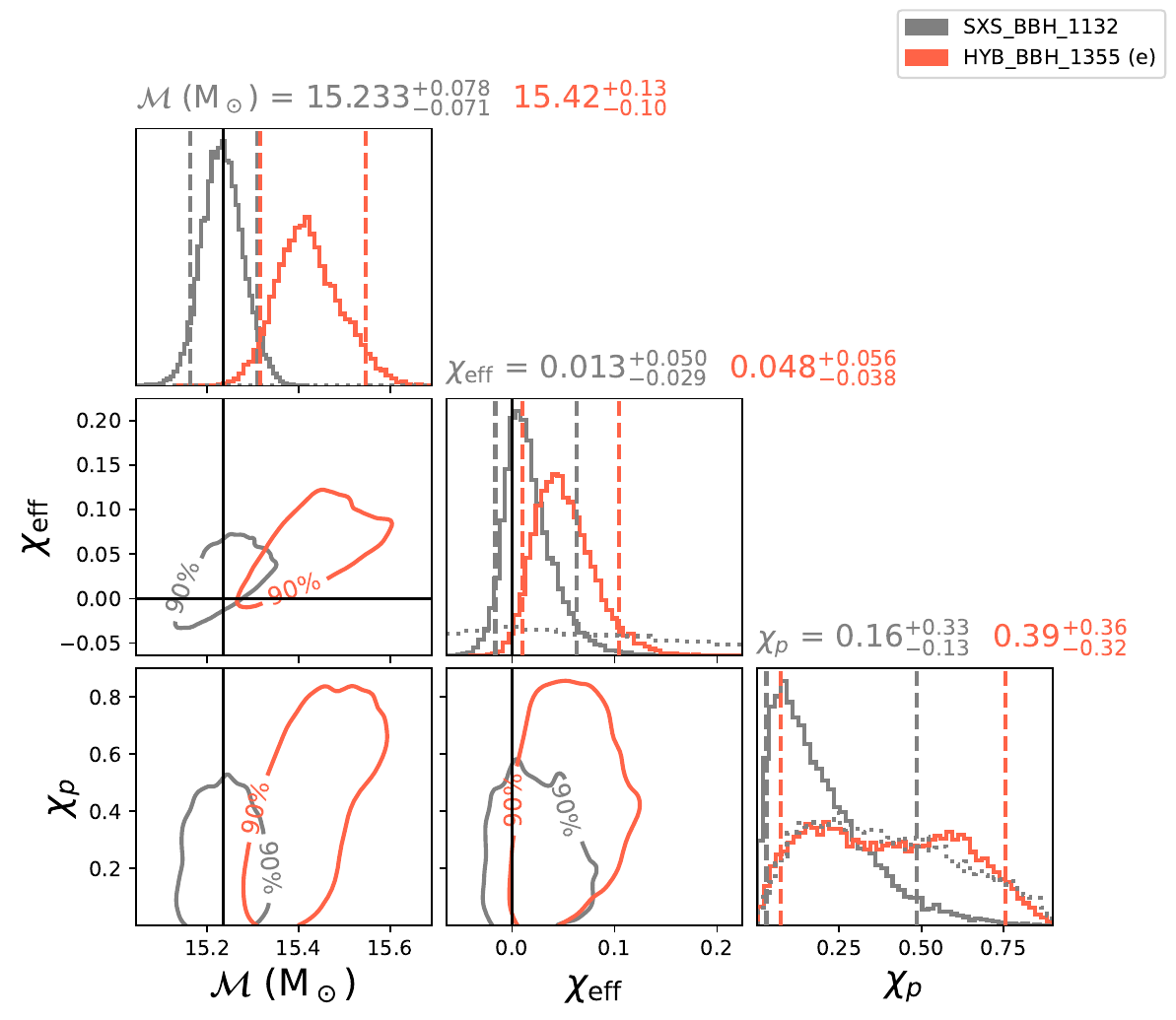}
    \includegraphics[trim=10 0 0 0, clip, width=0.49\linewidth]{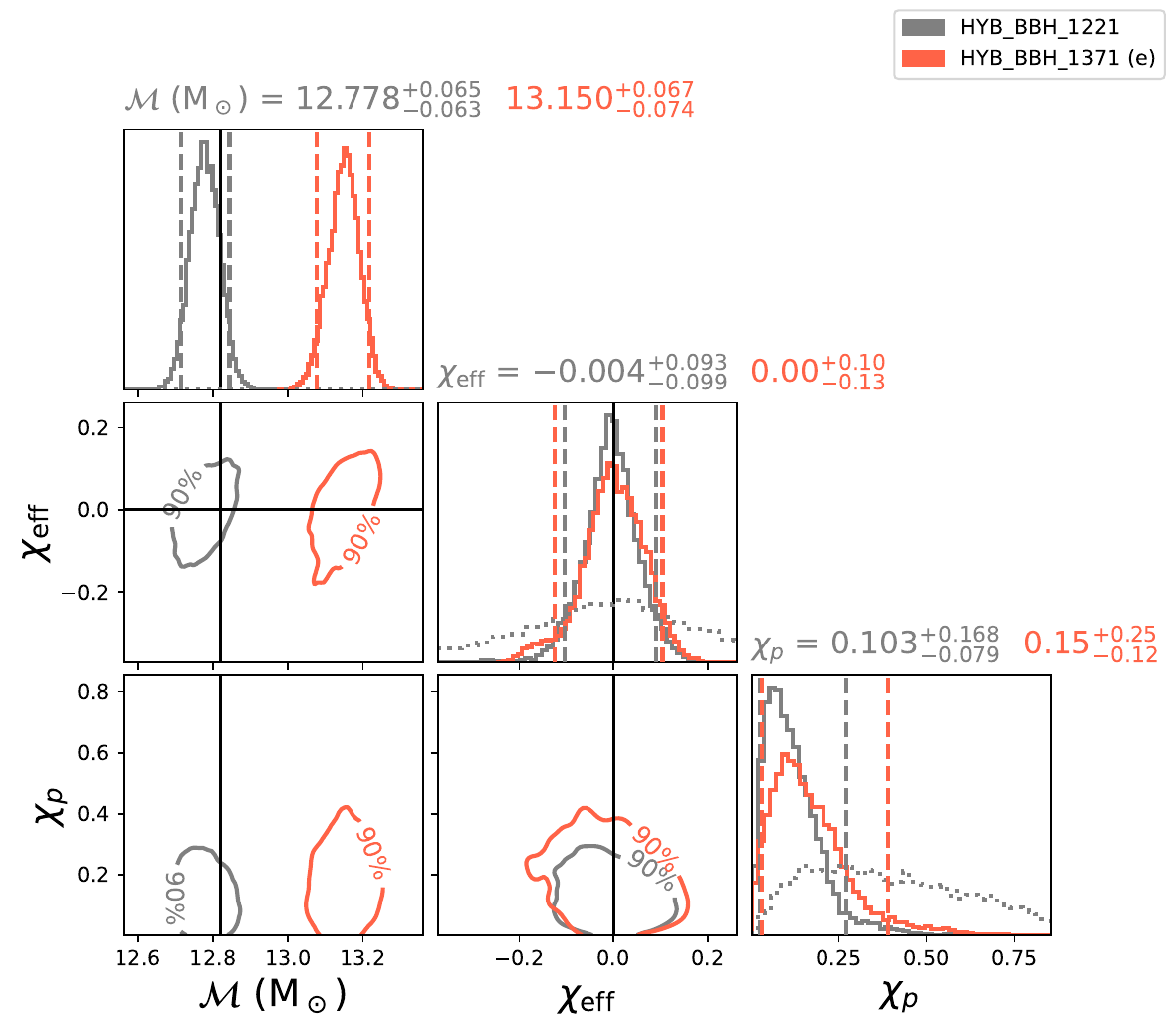}
    \caption{Same as Fig.~\ref{fig:corner-ps-q-2}, but for the cases of mass ratio $q=1$ (left) and $q=3$ (right).}
    \label{fig:corner-ps-q-1-3}
\end{figure*}

\begin{figure*}
    \centering
    \includegraphics[trim=10 0 0 0, clip, width=0.49\linewidth]{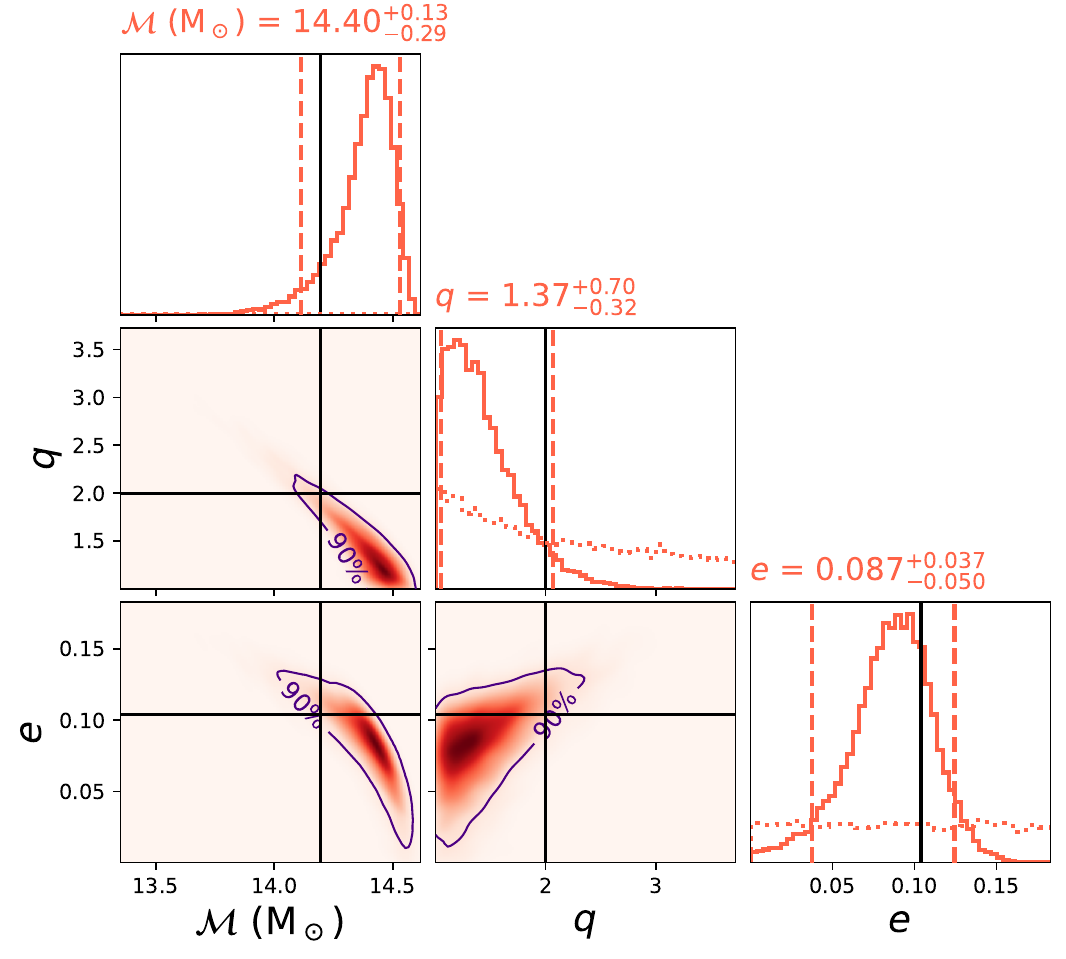}
    \includegraphics[trim=10 0 0 0, clip, width=0.49\linewidth]{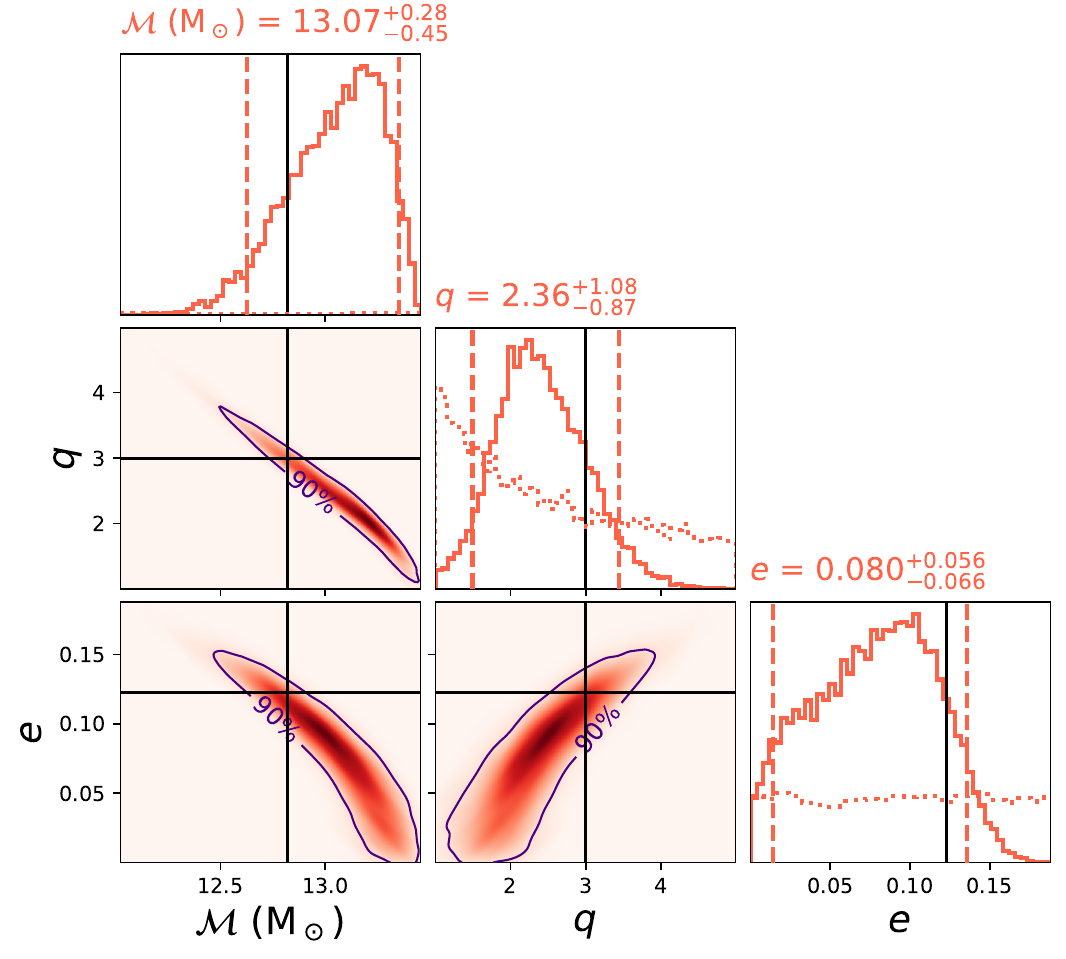}
    \caption{The corner plot, for $q=2$ (left) and $q=3$ (right), of chirp mass ($\mathcal{M}$), mass ratio ($q$), and eccentricity ($e_{20}$), for the non-spinning recovery from Fig.~\ref{fig:violin-ns-inj} performed using \textsc{TaylorF2Ecc}. The histograms on the diagonal of the plot are 1D marginalized posteriors for the respective parameters with vertical dashed lines denoting $90\%$ credible intervals. The dotted curves show the priors used.}
    \label{fig:corner-ns-ecc}
\end{figure*}

\begin{figure*}[p]
    \centering
    \includegraphics[trim=10 0 0 0, clip, width=0.49\linewidth]{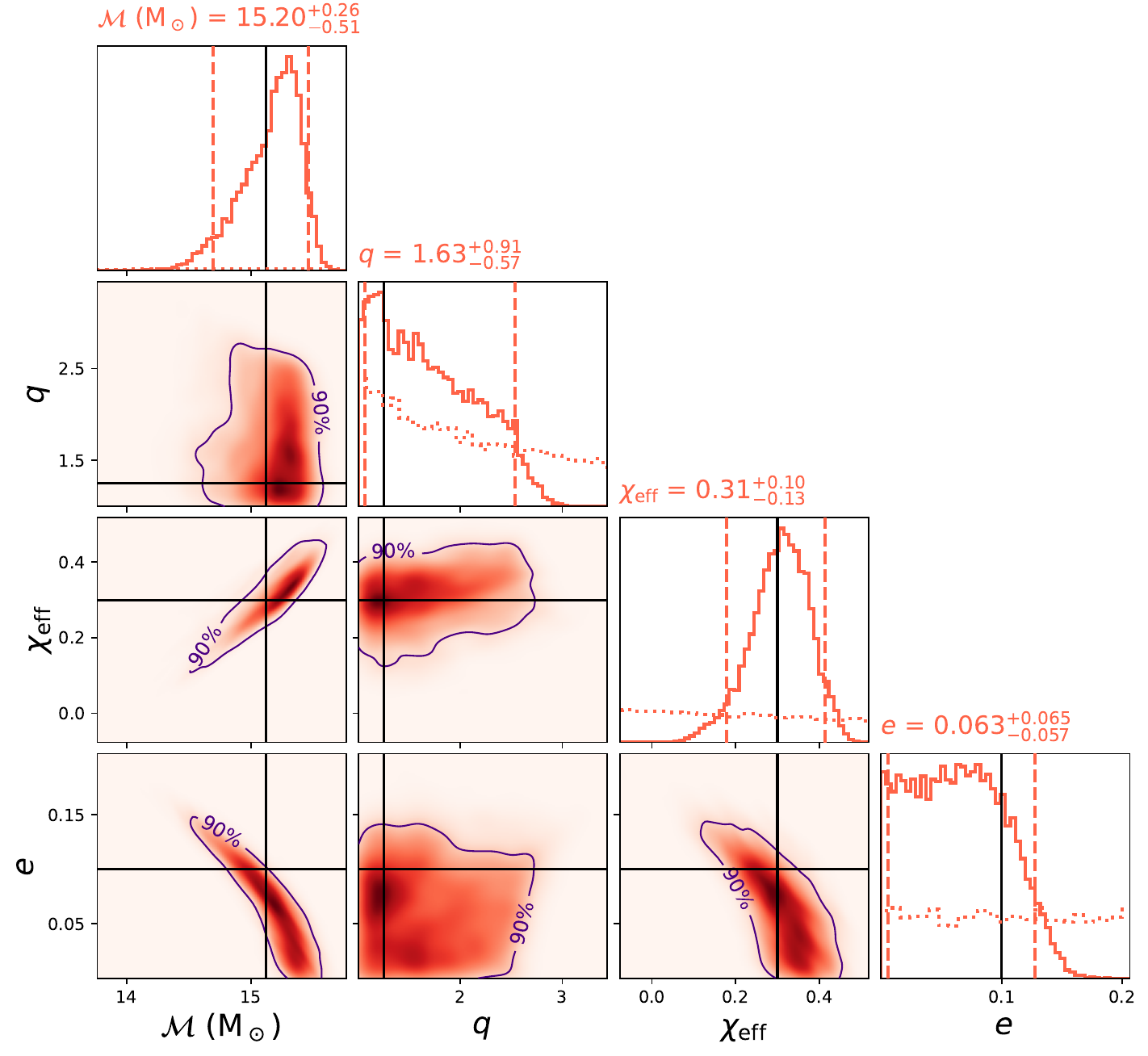}
    \includegraphics[trim=10 0 0 0, clip, width=0.49\linewidth]{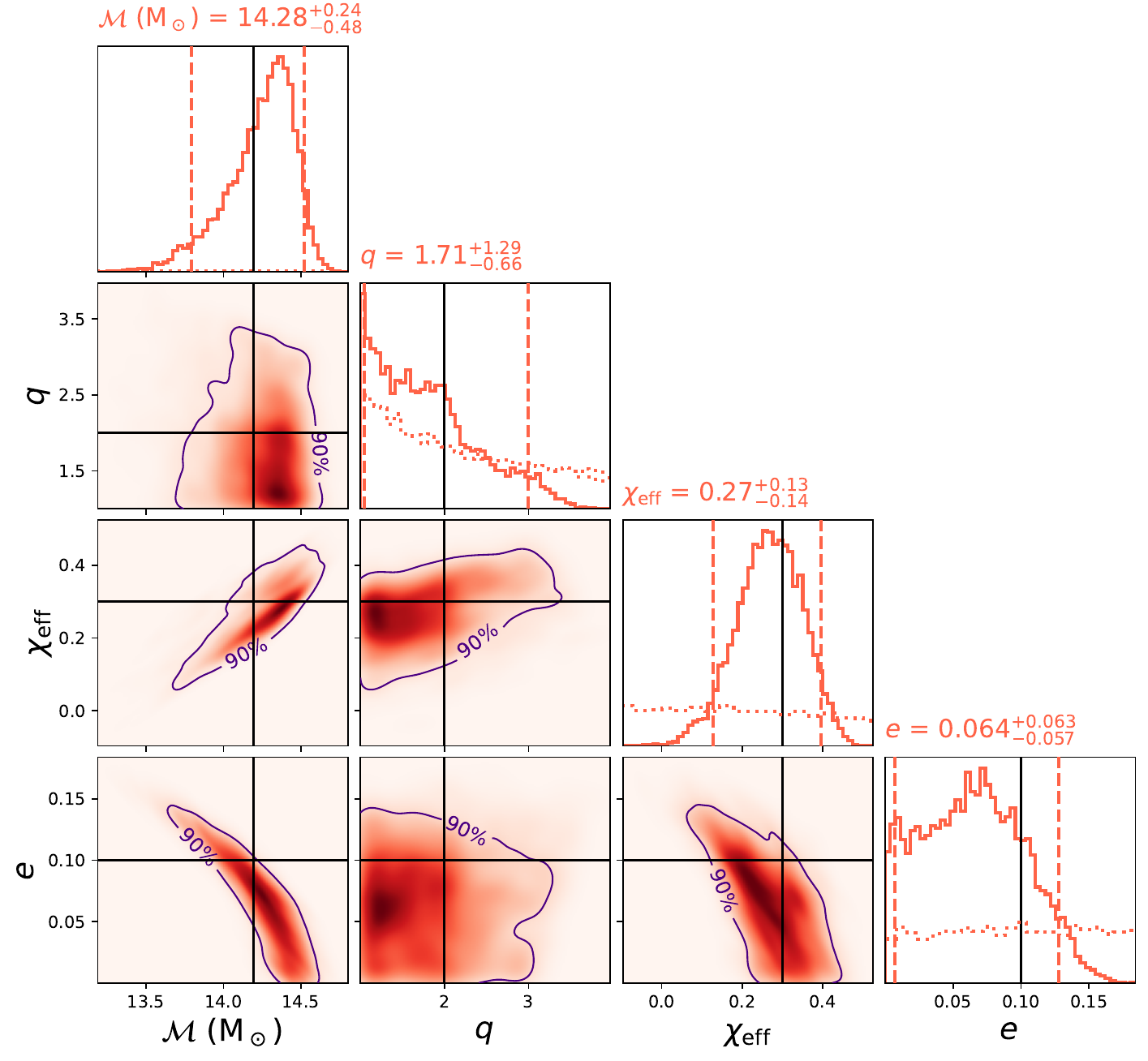}
    \includegraphics[trim=10 0 0 0, clip, width=0.49\linewidth]{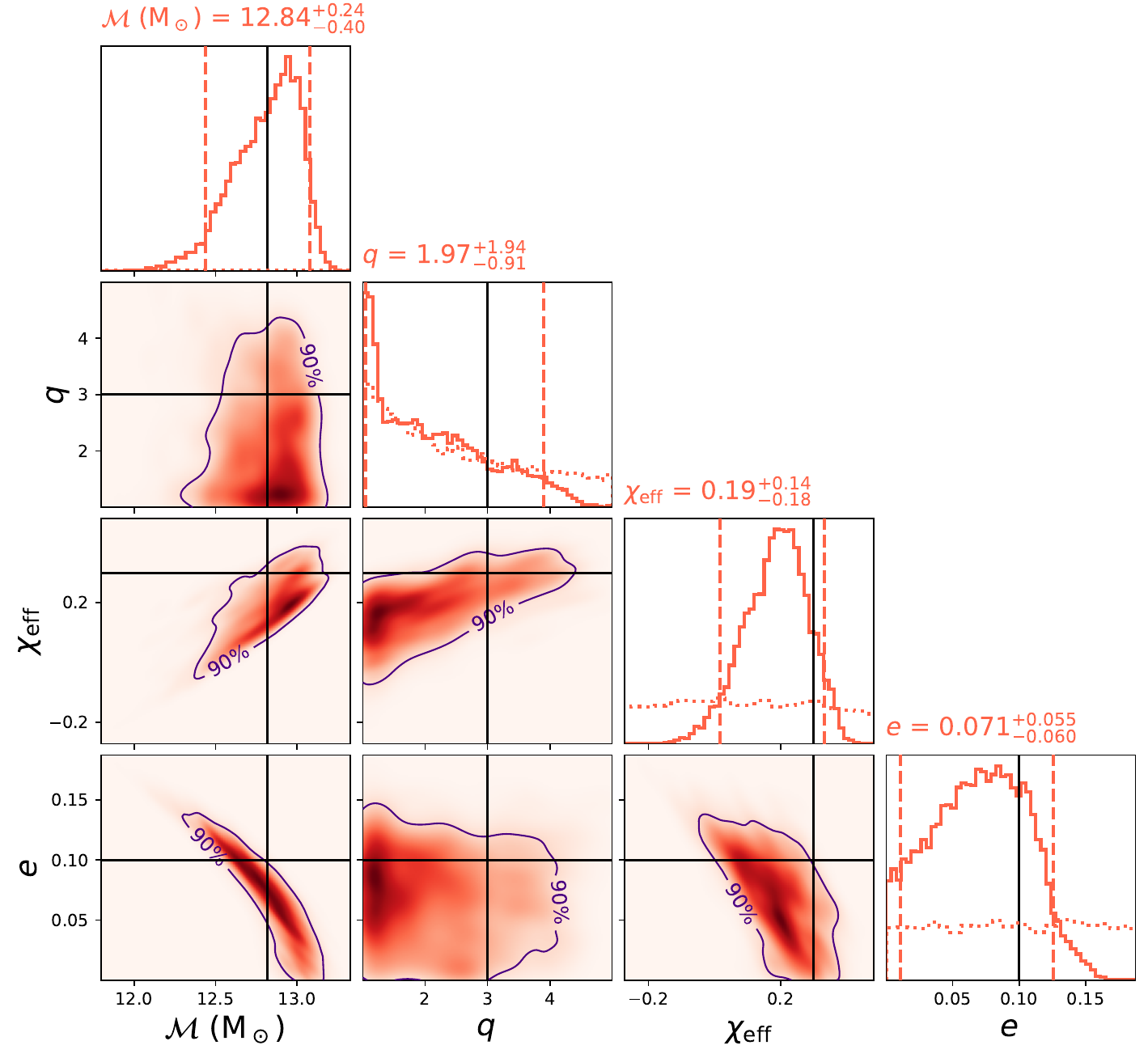}
    \caption{The corner plot, for mass ratios $q=1.25$ (top-left), $q=2$ (top-right) and $q=3$ (bottom), of chirp mass ($\mathcal{M}$), mass ratio ($q$), effective spin parameter ($\chi_\text{eff}$), and eccentricity ($e$), for the aligned spin recovery from Fig.~\ref{fig:violin-as-inj} performed using \textsc{TaylorF2Ecc}. The histograms on the diagonal of the plot are 1D marginalized posteriors for the respective parameters with vertical dashed lines denoting $90\%$ credible intervals. The dotted curves show the priors used.}
    \label{fig:corner-as-ecc}
\end{figure*}

\begin{figure*}
    \centering
    \includegraphics[trim=10 0 0 0, clip, width=0.45\linewidth]{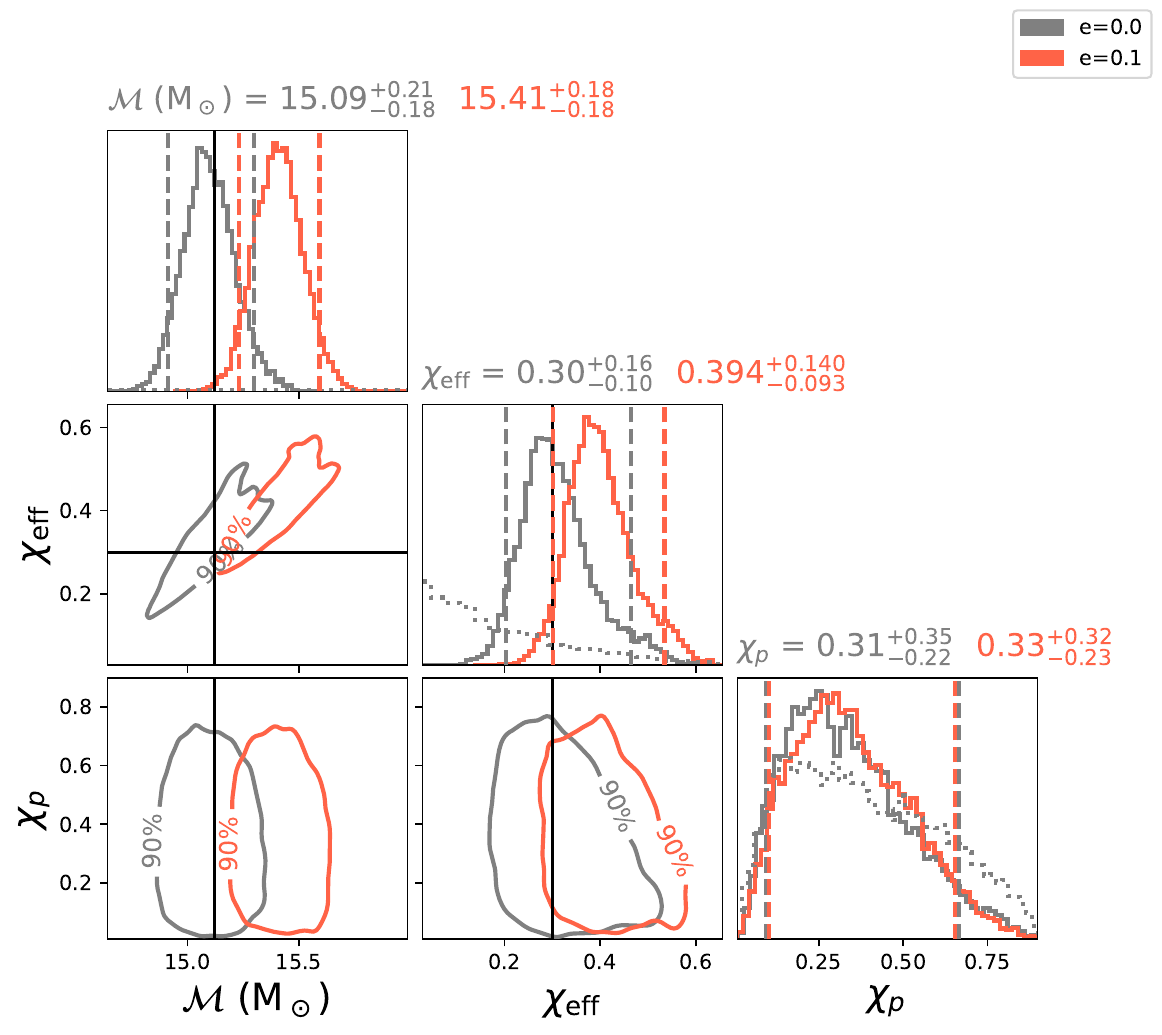}
    \includegraphics[trim=10 0 0 0, clip, width=0.45\linewidth]{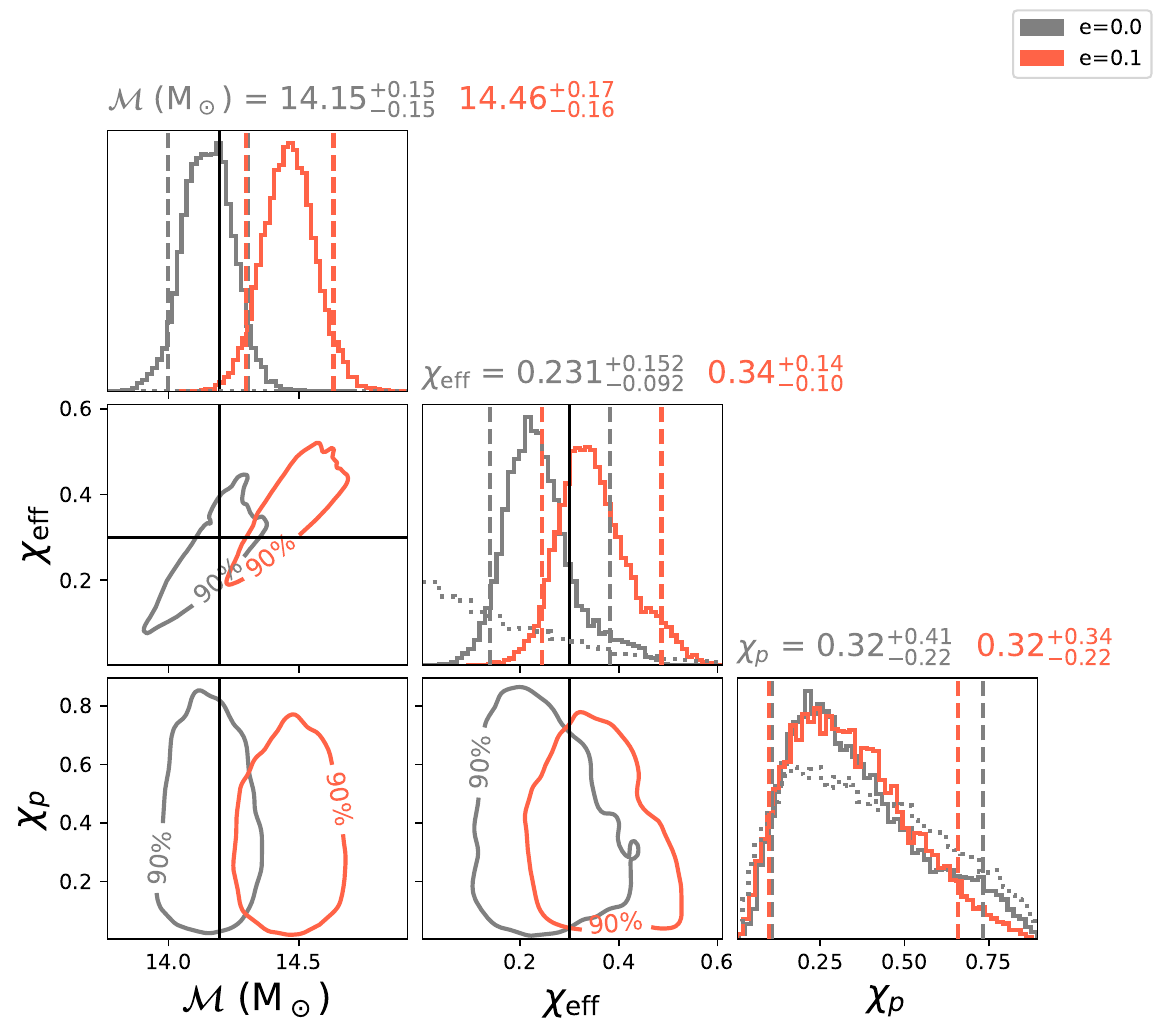}
    \includegraphics[trim=10 0 0 0, clip, width=0.45\linewidth]{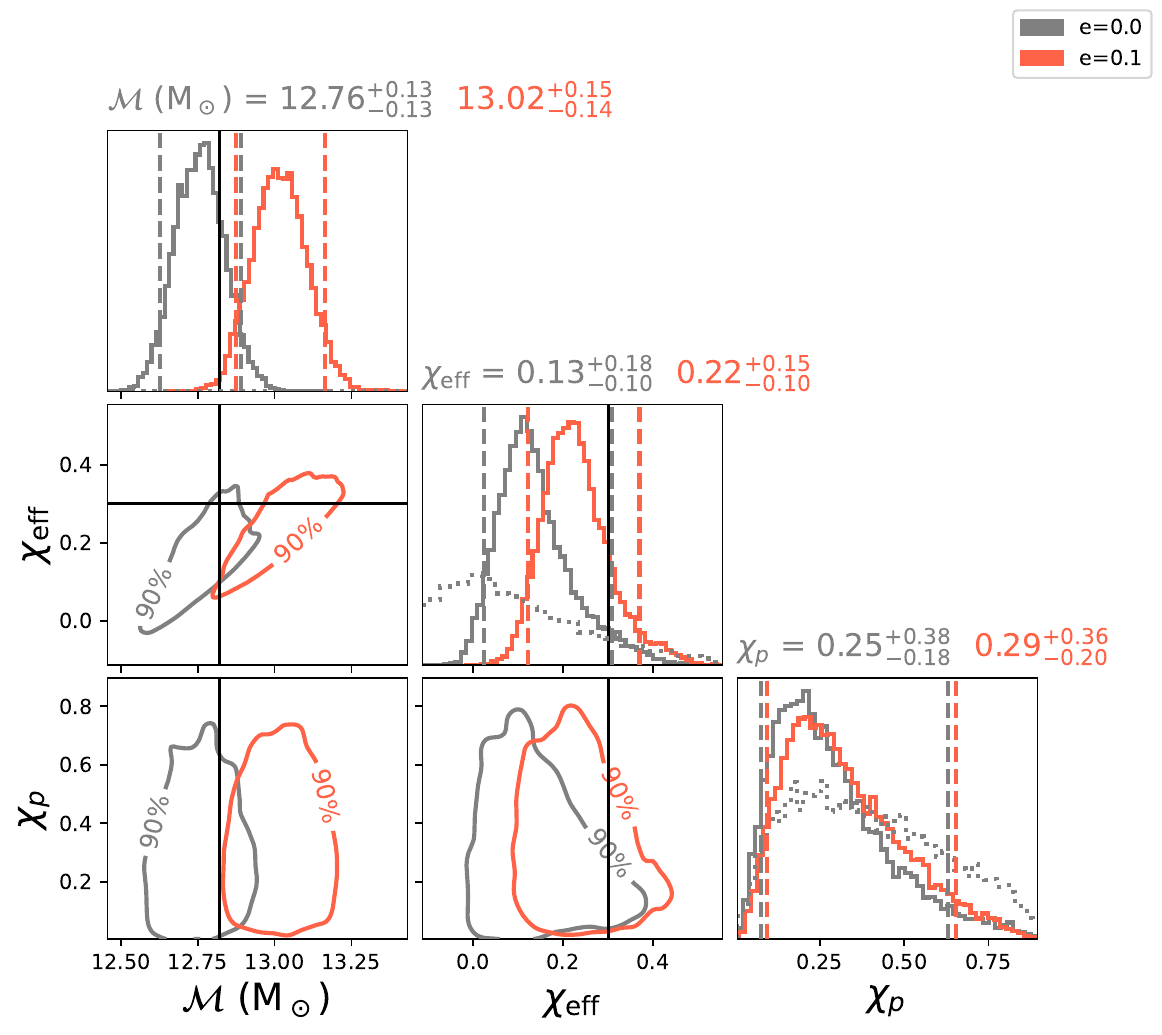}
    \caption{Same as Fig.~\ref{fig:corner-ps-q-2}, but for aligned-spin injections with mass ratios $q=1.25$ (top-left), $q=2$ (top-right), and $q=3$ (bottom).}
    \label{fig:corner-ps-q-2-3-as-inj}
\end{figure*}

\section{Note on equal-mass case ($q=1$)}
\label{appendix:q_1}

Here we discuss the $q=1$ case, which is different from the other mass ratio cases due to physical limits on the mass ratio prior ($q \geq 1$). The shift in the chirp mass posteriors for different spin configurations, seen in right panel of Fig.~\ref{fig:hist-ns-as-ps}, can partly be explained due to the prior railing of mass ratio ($q$) leading to a prior railing in component masses. Since the true value of injection is exactly $q=1$, parameters correlated with $q$ can become biased due to the entire posterior volume existing above the $q=1$ boundary. In order to confirm this, we carry out an identical baseline injection run where we inject ($\ell$=$2$, $m$=$|2|$) mode, non-spinning, quasi-circular signal into zero noise using \textsc{IMRPhenomXAS}, and recover it in the same three spin configurations (non-spinning, aligned spin, precessing spin) as used for the hybrids. We observe that the trends are identical to the ones observed using the hybrids. Hence we conclude that for $q=1$ case, the slight deviation from the usual trend is because of the mass ratio prior skewing the chirp mass posteriors.

\section{Summary of correlations of eccentricity with other parameters}
\label{appendix:ecc_corner}

Here we show the bounds on $e_{20}$
obtained using \textsc{TaylorF2Ecc} to recover injections. In Fig.~\ref{fig:corner-ns-ecc}, we show the posteriors on eccentricity, chirp mass and mass ratio. For both the $q=2$ and $q=3$ cases, the $90\%$ bounds on $e_{20}$ include the injected value. Looking at the 2D plots, we see that eccentricity shows negative and positive correlations with the chirp mass and mass ratio parameters, respectively. Further, when we look at the aligned-spin injections shown in figure \ref{fig:corner-as-ecc}, there is also a mild correlation with the effective spin parameter $\chi_\text{eff}$. Here again we see that the $90\%$ credible interval for the eccentricity parameter includes the injected value.

\section{Simulated noise injections: non-spinning and eccentric}
\label{appendix:noisy_injs}

\begin{figure*}
    \centering
    \includegraphics[trim=40 0 30 20, clip, width=0.32\linewidth]{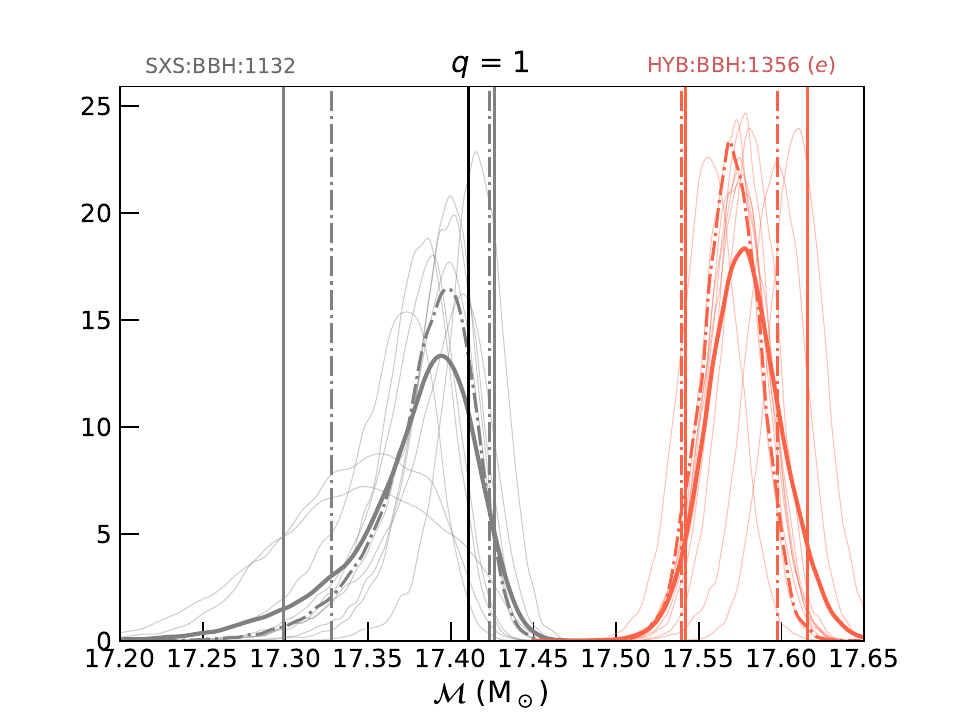}
    \includegraphics[trim=40 0 30 20, clip, width=0.32\linewidth]{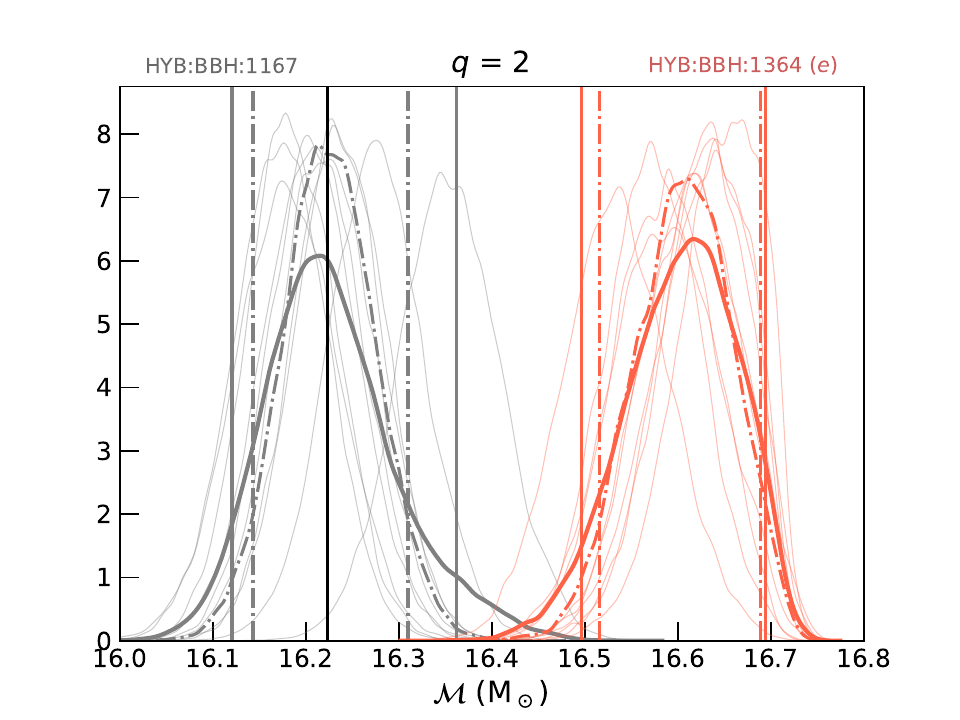}
    \includegraphics[trim=40 0 30 20, clip, width=0.32\linewidth]{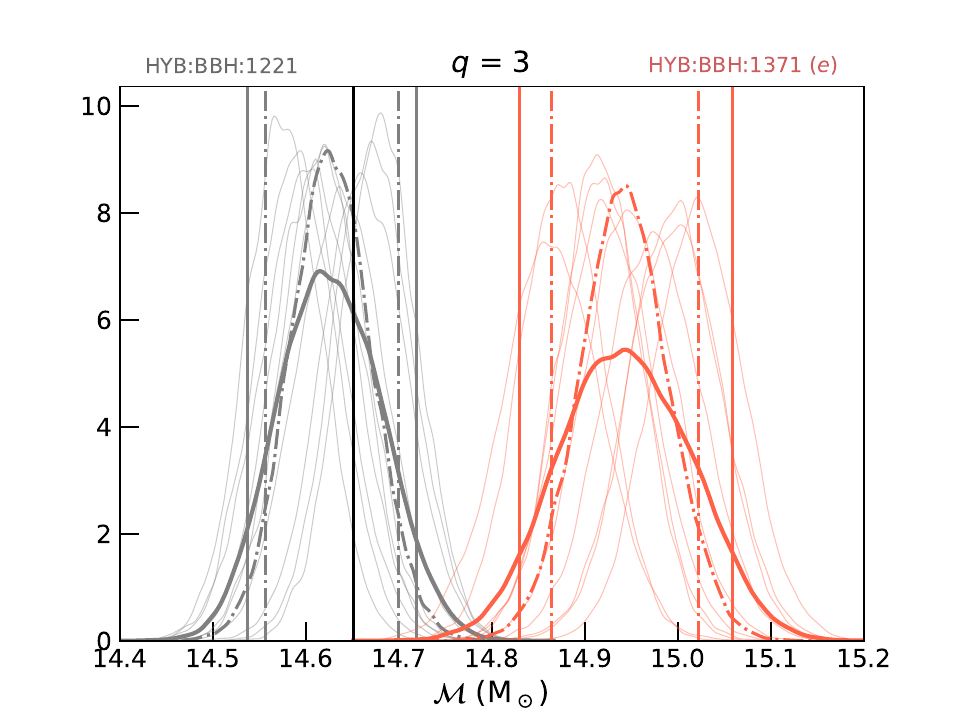}
    \caption{The chirp mass posteriors, for injections with mass ratios $q=1,2,3$, compared between eccentric (red) and quasi-circular (grey) non-spinning injections, all recovered using \textsc{IMRPhenomXAS} with zero spins. The faint lines depict injections in different noise realisations, and the dark solid curve is the combined posterior. We have also plotted the zero-noise injection as dashed curve. The coloured vertical lines represent the $90\%$ credible interval of the combined posterior of the same colour as the line, and the black line denotes the injected value.}
    \label{fig:noise-inj}
\end{figure*}

We perform a set of injection recoveries with Gaussian noise simulated using the power spectral density (PSDs) of the detectors. The total mass of the injected BBH systems is $40 M_\odot$ and the mass ratios are $q=1, 2, 3$, with the slightly heavier mass chosen to increase the computational efficiency of the analysis. These are recovered using quasi-circular waveform \textsc{IMRPhenomXAS} with zero spins. The results can be seen in Fig.~\ref{fig:noise-inj}. For each case, we have taken $10$ noise realizations, which each correspond to the posteriors shown by thin curves in the plot. An equal number of samples were taken from each of these runs and combined to form the average posterior shown by the thick coloured curve. We also perform a zero-noise injection for each mass ratio case, which is shown by the dot-dashed curve in the plot. The vertical coloured lines denote $90\%$ credible interval and the black line shows the injected value. As can be seen, the average posterior of all the noisy injections agrees well with the zero-noise curve for each case. Hence, for the analyses in the main text, we have only performed zero-noise injections.

\clearpage

%%%%%%%%%%%%%%%

\bibliographystyle{apsrev4-1}
\bibliography{master_refs}
%%%
\end{document}